\documentclass[prx,amsfonts,floatfix,superscriptaddress,longbibliography,twocolumn]{revtex4-1}
\usepackage{mathbbol}
\usepackage{graphicx,color}
\usepackage{amsmath,amssymb,bm,amsthm}
\usepackage{epstopdf}

\usepackage{bbding}

\def\be{\begin{equation}}
\def\ee{\end{equation}}
\def\ba{\begin{eqnarray}}
\def\ea{\end{eqnarray}}

\begin{document}
\title{Self-averaging in many-body quantum systems out of equilibrium: Chaotic systems}

\author{Mauro Schiulaz}
\address{Department of Physics, Yeshiva University, New York City, New York, 10016, USA}
\author{E. Jonathan Torres-Herrera}
\address{Instituto de F\'isica, Benem\'erita Universidad Aut\'onoma de Puebla,
Apt. Postal J-48, Puebla, 72570, Mexico}
\author{Francisco P\'erez-Bernal}
\address{Dep. CC. Integradas y Centro de Estudios Avanzados en F\'isica,
Matem\'aticas y Computaci\'on. Fac. CC. Experimentales, Universidad de Huelva, Huelva 21071, \& Instituto Carlos I de F\'isica Te\'orica y Computacional, Universidad de Granada, Granada 18071, Spain}
\author{Lea F. Santos}
\address{Department of Physics, Yeshiva University, New York City, New York, 10016, USA}

\date{\today}

\begin{abstract}
Despite its importance to experiments, numerical simulations, and the development of theoretical models, self-averaging in many-body quantum systems out of equilibrium remains underinvestigated. Usually, in the chaotic regime, self-averaging is taken for granted. The numerical and analytical results presented here force us to rethink these expectations. They demonstrate that self-averaging properties depend on the quantity and also on the time scale considered. We show analytically that the survival probability in chaotic systems is not self-averaging at any time scale, even when evolved under full random matrices. We also analyze the participation ratio, R\'enyi entropies, the spin autocorrelation function from experiments with cold atoms, and the connected spin-spin correlation function from experiments with ion traps. We find that self-averaging holds at short times for the quantities that are local in space, while at long times, self-averaging applies for quantities that are local in time. Various behaviors are revealed at intermediate time scales.
\end{abstract}

\maketitle

\section{Introduction}

The property of self-averaging is at the heart of studies about disordered systems~\cite{LifshitzBook} and random matrices~\cite{MehtaBook}. It holds when the distribution of the quantity of interest is peaked around its average and its relative variance goes to zero as the system size increases. For sufficiently large systems, the distribution converges to a delta function. This implies that, by increasing the system size, one can reduce the number of samples used in experiments and in statistical analysis. If the system exhibits self-averaging, its physical properties are independent of the specific realization, which allows for the construction of theoretical models to describe finite samples.  Lack of self-averaging, on the other hand, means sample to sample fluctuations even in the thermodynamic limit, so ensemble averages are needed no matter how large the system size is. In this case, scaling analyses become quite challenging.

Absence of self-averaging typically happens near critical points of disordered systems~\cite{Wiseman1995,Aharony1996,Wiseman1998,Castellani2005,Malakis2006,Roy2006,Monthus2006,Efrat2014}. This is the case of one-body~\cite{MullerARXIV} and many-body~\cite{Serbyn2017} systems in the vicinity of the metal-insulator transition. Using results from Anderson localization, it has been shown, for example, that the entanglement entropy is not self-averaging~\cite{Pastur2014PRL}. Self-averaging has also been the theme of works about spin glass~\cite{Castellani2005,Wreszinski2004}, the kinetics of domain growth~\cite{Milchev1986}, and diffusion in disordered media~\cite{Bouchaud1990,AkimotoPRL2016,Russian2017,AkimotoPRE2018}, often in comparison to ergodicity.

Ergodicity refers to temporal averages~\cite{Mithun2018} and has recently received extensive attention in studies about equilibration and thermalization of isolated quantum systems~\cite{Deutsch1991,SrednickiARXIV,Rigol2008,Dalessio2016,Reimann2008,Short2011,Short2012,Zangara2013,Nation2018}. Self-averaging, on the other hand, is associated with averages over disorder realizations and it has got little consideration in the context of quantum systems out of equilibrium, apart from few recent works on driven systems~\cite{Lobejko2018,Mukherjee2018} and studies about the two-level form factor~\cite{Leviandier1986,Argaman1993b,Eckhardt1995,Prange1997,Braun2015}, which is a quantity used to study spectral properties in the time domain.

The present work addresses the mostly uncharted territory of self-averaging during the evolution of interacting many-body quantum systems. The focus is on the chaotic regime, where self-averaging is usually assumed to hold. Our results invalidate these expectations. 

We show analytically and confirm numerically that the survival probability evolving under full random matrices from the Gaussian orthogonal ensemble (GOE) is not self-averaging at any time scale.  We also study the survival probability in a chaotic disordered spin model of experimental interest and verify that it is nowhere self-averaging. This is a consequential result, since this spatially non-local quantity is a very common tool in studies of nonequilibrium quantum dynamics~\cite{Fonda1978,Bhattacharyya1983,Ufink1993,Wilkie1991,Alhassid1992,Heller1991,Ufink1993,Khalfin1958,MugaBook,Campo2016NJP,Tavora2016,Tavora2017,Torres2014PRA,Torres2014NJP,TorresKollmar2015,Mazza2016,VolyaARXIV,Heyl2013,Santos2015,Ketzmerick1992,Torres2015,Bera2018,Lerma2018,Reimann2016,Reimann2019}. As we show, if self-averaging is assumed and one decreases the number of random realizations used as the system size increases, one misses essential features of the evolution of the survival probability. The same is expected for other non-self-averaging quantities.

Our analysis is extended also to other non-local and local quantities evolved with both the GOE and the disordered spin model. As examples of non-local quantities, we consider the participation ratio and R\'enyi entropies. They measure the spread of the initial state in the many-body Hilbert space and are connected with the out-of-time ordered correlator~\cite{Fan2017,Borgonovi2019b}. As local observables, we investigate the spin autocorrelation function, which is similar to the density imbalance measured in experiments with cold atoms~\cite{Schreiber2015}, and the connected spin-spin correlation function, which is used in experiments with ion traps~\cite{Richerme2014}. The results are rather non-trivial, being dependent on the observable and time scales, although a general picture emerges for short and long times.

The two spatially local quantities are self-averaging at the short times currently accessible experimentally, which is reassuring. After equilibration, when there are only small fluctuations around infinite-time averages, the connected spin-spin correlation function, the participation ratio, and the R\'enyi entropies are all self-averaging, while the survival probability and the spin autocorrelation function are not. These two latter quantities are autocorrelation functions, being therefore non-local in time. 

This paper is organized as follows. The definition of self-averaging, as well as the  models, initial states, and quantities investigated are presented in Sec.~\ref{Sec:Def}. Section~\ref{Sec:mean} shows the entire evolution of the average values of the observables under the GOE and the spin model. The plots display the different time scales involved in the dynamics and they serve as references for the following core sections. Section~\ref{Sec:SP} provides the analytical and numerical results for the survival probability. The other non-local quantities are studied in Sec.~\ref{Sec:IPR}, and the local experimental observables are examined in Sec.~\ref{Sec:Imb}. An explicit example of the consequences one may face when decreasing the number of random realizations for a quantity that is non-self-averaging is given in Sec.~\ref{Sec:Lack}. Conclusions and future directions are presented in Sec.~\ref{Sec:Conc}.  There are also two appendices. 

\section{General definitions}
\label{Sec:Def}

In this section, we define the concept of self-averaging, and introduce the Hamiltonians and quantities examined in this work.

\subsection{Self-averaging}

Self-averaging implies that a single large system is enough to represent the whole statistical ensemble. By analyzing the sample to sample fluctuations, a quantity $O$ is said to be self-averaging when the ratio between its variance $\sigma^2_{O}$ and the square of its mean, that is, its relative variance~\cite{Wiseman1995,Aharony1996,Wiseman1998,Castellani2005,Malakis2006,Roy2006,Monthus2006,Efrat2014,Lobejko2018},
\be
{\cal R}_{O}(t)=\frac{\sigma^2_{O}(t)}{\left<O(t)\right>^2} = \frac{\left<O^2(t)\right>-\left<O(t)\right>^2}{\left<O(t)\right>^2} ,
\label{eq:sigma} 
\ee
goes to zero as the system size $L$ increases. In our studies, $\langle.\rangle$ includes the averages over both disorder realizations and initial states taken in a narrow energy window around the middle of the spectrum. Notice that our ${\cal R}_{O}(t)$ is time-dependent, since we investigate whether the observable is self-averaging not only at equilibrium, but during its entire time evolution.

It is common to distinguish strong self-averaging, when ${\cal R}_{O}(t) \sim L^{-1}$, from weak self-averaging, when ${\cal R}_{O}(t) \sim L^{-\nu}$ for  $0<\nu<1$. In this work, we find also more extreme cases, in the sense that the relative variance of $O$ can decrease or increase exponentially in system size. This sort of ``super" self-averaging or ``super" non-self-averaging behavior occurs at large times, when the dynamics of a chaotic many-body system become analogous to those of full random matrices~\cite{Schiulaz2019}. At such long times, the initial state is spread over the many-body Hilbert space, which is exponentially large in $L$.

Equation (\ref{eq:sigma}) is the standard definition of self-averaging~\cite{Wiseman1995,Aharony1996,Wiseman1998,Castellani2005,Malakis2006,Roy2006,Monthus2006,Efrat2014,Lobejko2018}. One may, of course, have observables for which $\left<O(t)\right>^2$ goes to zero at long times as the system size increases, which is indeed the case of all quantities studied here, except for the entropies. The question we address is whether the variance goes to zero faster than $\left<O(t)\right>^2$.

\subsection{Models and Initial States}

We study quantum Hamiltonians of the form
\be
H=H_0+ V,
\label{eq:H}
\ee
where $H_0$ is the integrable part of $H$ and $V$ is a strong perturbation that takes the system into the chaotic regime. We denote by $|n\rangle$ the eigenstates of $H_0$. The eigenstates and eigenvalues of $H$ are  $|\alpha\rangle$ and $E_\alpha$, respectively.

\subsubsection{GOE model}
One of the models that we study is formed by GOE full random matrices of dimension $D$ \cite{MehtaBook}. For this model, $H_0$ corresponds to the diagonal part of the matrix, while $V$ contains the off-diagonal elements. All entries are real numbers independently drawn from a Gaussian distribution with mean value $\left<\left<H_{ij}\right>\right>=0$ and variance
\be
\left<\left<H_{ij}^2\right>\right>=\begin{cases}
2 & i=j \\
1 & i\neq j
\end{cases}.
\ee
This model is unphysical, because it assumes interactions between all degrees of freedom, but it has the advantage of allowing for analytical calculations. This is possible, because the eigenvalues of the GOE model are highly correlated and the eigenstates are normalized random vectors. This model is also relevant, because it correctly reproduces the spectral correlations and the late time dynamics of realistic models~\cite{Schiulaz2019}.

\subsubsection{Disordered spin model}

We consider a realistic disordered spin-$1/2$ chain in the strong chaotic regime. It has local two-body interactions only and its Hamiltonian is given by,
\ba
H_0 &=& J \sum_{k=1}^L (h_k S_k^z +  S_k^z S_{k+1}^z) \nonumber \\
V &=& J\sum_{k=1}^L (S_k^x S_{k+1}^x + S_k^y S_{k+1}^y).
\label{eq:Hspin}
\ea
In the above, $\hbar=1$, $S_k^{x,y,z}$ are spin operators on site $k$, $L$ is the number of spins in the lattice,  $J$ sets the energy scale, and periodic boundary conditions are taken. The Zeeman splittings $h_i$ are independent random numbers uniformly distributed in $[-h,h]$, where $h$ is the disorder strength. The total magnetization in the $z$-direction is conserved. We work in the largest subspace, namely the one with zero total $z$-magnetization, which has dimension $D=L!/(L/2)!^2$. To be in the fully chaotic region, we fix $h=0.75$. This model has been extensively studied in the context of many-body localization, both theoretically~\cite{SantosEscobar2004,Dukesz2009,Pal2010,Luitz2017} and experimentally~\cite{Schreiber2015}.

While, for the sake of concreteness, our calculations and numerical simulations are done for this model, they can be extended to a large class of different systems~\cite{Schiulaz2019}. The important elements that the model has to satisfy are to be strongly chaotic, in the sense of level statistics comparable to those of random matrix theory, and have interactions that are strictly local and two body only.

\subsubsection{Initial state and notation for ${\cal R}(t)$}

The initial state $\left|\Psi (0)\right>$ is an eigenstate of $H_0$, as often considered in experiments with cold atoms and ion traps. It has energy close to the middle of the spectrum,
\be
E_0 = \langle \Psi(0)|H|\Psi(0)\rangle = \sum_\alpha\left|c_\alpha^0\right|^2 E_\alpha \sim 0,
\ee 
where $c_\alpha^0=\left<\alpha|\Psi (0)\right>$. This is the region of the spectrum where the energy eigenstates are chaotic~\cite{Santos2012PRE}. Such initial states are very far from equilibrium, which results in an extremely fast initial evolution under the full Hamiltonian $H$.

In our numerical simulations, our averages are performed over $0.01D$  initial states with energy closest to the center of the spectrum and over many disorder realizations. The total amount of data considered in each average is $10^4$.

For clarity, we refer to the relative variance obtained for GOE matrices as ${\cal R}^\textrm{GOE}_O(t)$ and as ${\cal R}^\textrm{spin}_O(t)$ the one obtained for the chaotic spin model.

\subsection{Quantities}

We consider both non-local quantities and local experimental observables. 

\subsubsection{Survival Probability} 

The survival probability is the squared overlap between the initial state and its time evolved counterpart, 
\be
P_S(t)=\left|\left<\Psi (0)\right|e^{-iHt}\left|\Psi (0)\right>\right|^2 .
\label{eq:PS}
\ee
This quantity is non-local in space and also in time, since it compares the state at time $t$ with the state at time $t=0$. It has been studied in many different contexts, from the decay of unstable nuclei~\cite{Fonda1978}, the quantum speed limit~\cite{Bhattacharyya1983,Ufink1993}, and the onset of power-law decays~\cite{Khalfin1958,MugaBook,Campo2016NJP,Tavora2016,Tavora2017}, to quench dynamics~\cite{Torres2014PRA,Torres2014NJP,TorresKollmar2015,Mazza2016,VolyaARXIV}, ground state and excited state quantum phase transitions~\cite{Heyl2013,Santos2015}, and multifractality in one-body and many-body systems~\cite{Ketzmerick1992,Torres2015,Bera2018}. The survival probability is related to the two-level form factor studied in~\cite{Prange1997,Chan2018PRX,Bertini2018,SuntajsARXIV,FriedmanARXIV}, but this one contains information about the eigenvalues only, while the survival probability contains information about the initial state also, being therefore more appropriate for studies of dynamics. 

The survival probability can be written in the following useful integral representation,
\be
P_S(t)=\left| \int dE e^{-iEt} \rho_0(E) \right|^2,
\ee
where
\be
\rho_0(E)=\sum_{\alpha}\left|c_\alpha^0\right|^2\delta(E-E_\alpha)
\ee
is the energy distribution of the initial state, known as local density of states (LDOS). The width $\Gamma_0$ of the LDOS is related to the number of states $|n\rangle $  that are directly coupled to $\left|\Psi (0)\right>$ according to
\be
\Gamma_0^2 = \sum_{n \neq 0} |  \langle n |H| \Psi(0) \rangle |^2.
\label{eq:Gamma}
\ee
In the above, the sum runs over all states $\left|n\right>$, apart from $\left|\Psi(0)\right>$. For the GOE model, the average over initial states and disorder realizations naturally gives 
\be
\left<\Gamma_0^2\right>^{\text{GOE}}=D,
\label{Eq:GammaGOE}
\ee
while the sparsity of the spin Hamiltonian implies that 
\be
\left<\Gamma_0^2\right>^{\text{spin}} = \frac{J^2 L^2}{8 (L-1)} \sim  \frac{J^2 }{8 } L .
\label{Eq:GammaSpin}
\ee
This difference has important consequences for the time scales involved in the evolution of the mean value of the observables and in their self-averaging behavior. In what follows, we use the notation $\Gamma = \sqrt{\left<\Gamma_0^2\right>}$.

\subsubsection{Inverse Participation Ratio and R\'enyi Entropies} 

The inverse participation ratio and the R\'enyi entropies are non-local quantities in space, but they are local in time. In contrast to the survival probability, they compare the state at time $t$ with all states $|n\rangle$, not only with $| \Psi(0) \rangle$. 

The  inverse participation ratio is defined as
\be
\text{IPR}(t)=\sum_n\left|\left<n\right|e^{-iHt}\left|\Psi (0)\right>\right|^4 .
\label{eq:IPR}
\ee
It quantifies the spread of the initial many-body state in the basis of unperturbed many-body states  $\left|n\right>$ \cite{Borgonovi2019}.  At $t=0$, when the initial state is fully localized in this basis, $\text{IPR}(0)=1$. A state completely delocalized at time $t$ has $\text{IPR}(t)\sim 1/D$.

The second-order R\'enyi entropy is related to the inverse participation ratio as
\be
S(t)=-\ln [ \text{IPR}(t)] ,
\label{eq:Sr}
\ee
where no partial trace of degrees of freedom is involved. While the asymptotic value of $\text{IPR}(t)$ scales with the inverse of the size of the exponentially large Hilbert space, the maximum value of $S(t)$ is proportional to $L$. The motivation to study not only $\text{IPR}(t)$, but also its logarithm, comes from the knowledge that the logarithm cuts the tails of the distribution, therefore enhancing self-averaging properties. 

The results for $S(t)$ are equivalent to those for the Shannon entropy (or first-order R\'enyi entropy), which is written as
\be
Sh (t) = - \sum_n \left|\left<n\right|e^{-iHt}\left|\Psi (0)\right>\right|^2 \ln \left|\left<n\right|e^{-iHt}\left|\Psi (0)\right>\right|^2 .
\label{eq:Sh}
\ee
This entropy is often used in studies of quantum chaos (see~\cite{Torres2016,Torres2017Philo,Arranz2019} and references therein).

\subsubsection{Spin Autocorrelation Function\\ and Connected Correlation Function} 

The spin autocorrelation function and the connected spin-spin correlation function are experimental quantities. They are both local in space, but only the latter is also local in time. 

The spin autocorrelation function  is given by
\be
I(t)=\frac{4}{L}\sum_{k=1}^L \left<\Psi (0) \right|S^z_k e^{iHt} S^z_k e^{-iHt}\left|\Psi (0) \right>.
\label{eq:I}
\ee
It measures the average over all sites of the proximity of the orientation of a spin $k$ at time $t$ to its orientation at $t=0$. By mapping the spin system to hardcore bosons, one finds that this quantity is analogous to the density imbalance between even and odd sites, which is measured in cold atom experiments~\cite{Schreiber2015}.

The connected spin-spin correlation function is defined as
\ba
C(t)&=&\frac{4}{L}\sum_k\left[\left<\Psi(t)\right|S_k^zS_{k+1}^z\left|\Psi(t)\right>\right.\\
&-&\left.\left<\Psi(t)\right|S_k^z\left|\Psi(t)\right>\left<\Psi(t)\right|S_{k+1}^z\left|\Psi(t)\right>\right].\nonumber
\ea
Similar to $\text{IPR}(t)$, it quantifies how far $|\Psi(t)\rangle$ is from the classical states $|n\rangle$. This quantity has been measured in experiments with ion traps~\cite{Richerme2014}. 

\section{Dynamics of mean values}
\label{Sec:mean}

Before studying in detail the behavior of the relative variance of the quantities above, we briefly explain how the average values, $\left<O(t)\right>$, change with time. We outline the main stages of the evolution and the time scales associated with these steps, so that in the following sections, we can analyze how the fluctuations behave in each of these regions.

\begin{figure}[h!]
\hspace{-0.78 cm}
\includegraphics*[width=0.47\textwidth]{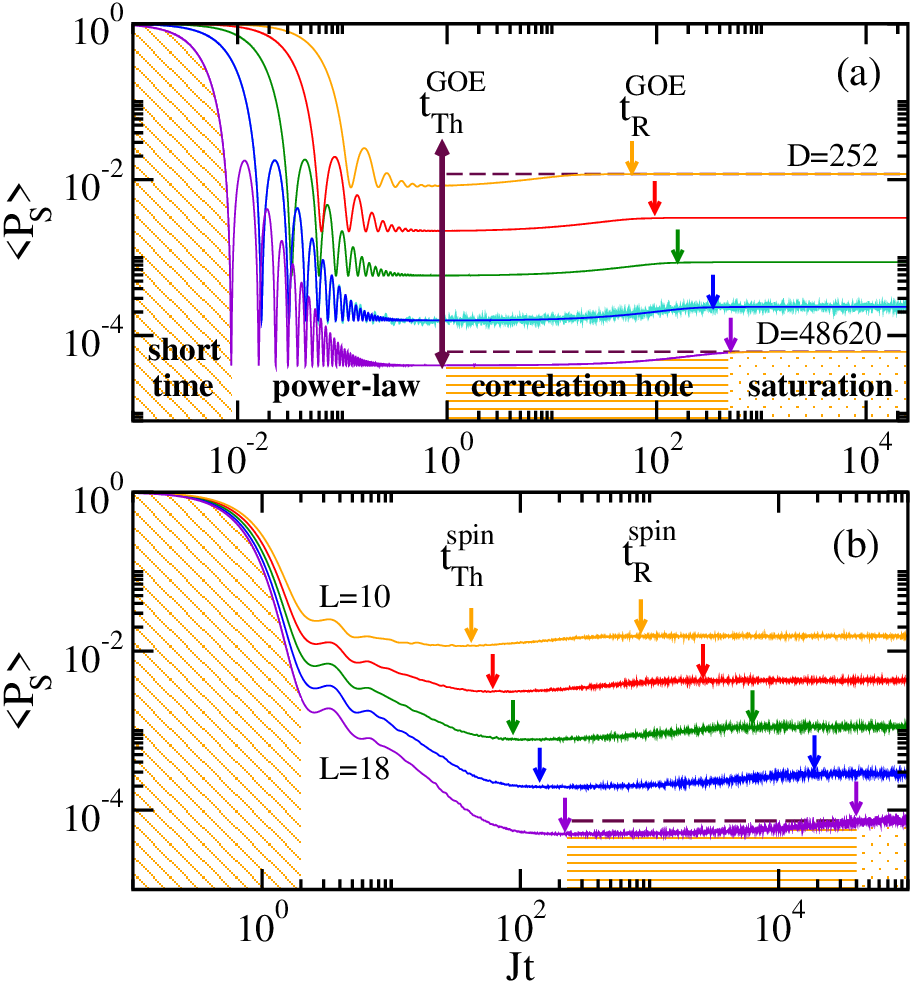}\\
\hspace{0.28 cm}
\hspace{-0.78 cm}
\includegraphics*[width=0.45\textwidth]{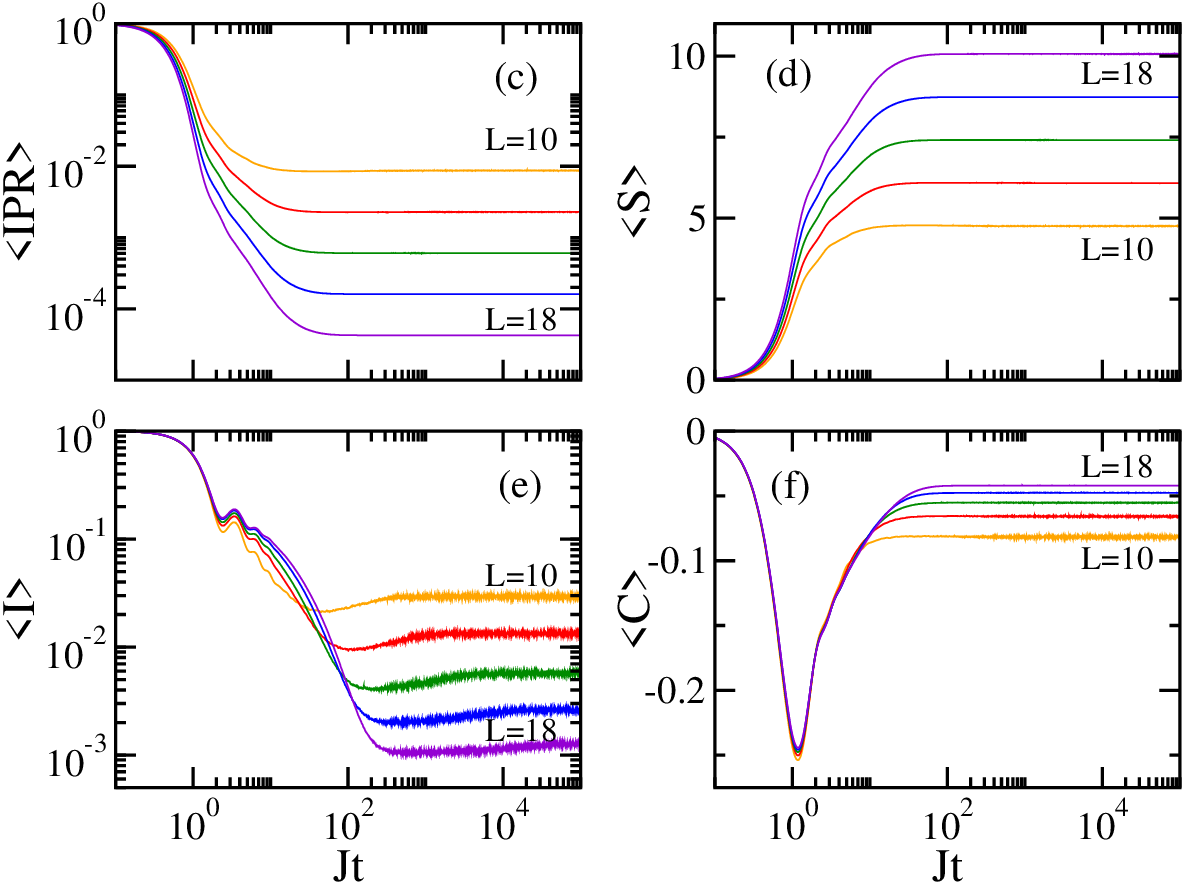}
\caption{Evolution of the mean value of the survival probability for the GOE model (a) and for the chaotic spin model (b), and evolution of the mean value of the inverse participation ratio (c), second-order R\'enyi entropy (d), spin autocorrelation function (e), and connected spin-spin correlation function (f)  for the chaotic spin model. The system sizes are indicated in the panels. In (a)-(c) and (e), from the top curve to the bottom curve: $D=252,924,3\,432,12\,870,48\,620$ (orange, red, green, blue, purple); in (d) and (f), these sizes are from the bottom curve to the top curve. In (a) and (b): Horizontal dashed lines mark the saturation value. In (a): Analytical expression from Eq.~(\ref{Eq:PsGOE}) and numerical data for $D=12\,870$. In (b-(f)): Numerical data. All panels: Average over $10^4$ data, where $0.01D$ different initial states with $E_0\sim0$ are selected for each disorder realization.}
\label{fig:PsGOE}
\end{figure}

In Fig.~\ref{fig:PsGOE}~(a) and Fig.~\ref{fig:PsGOE}~(b), we show the evolution of the mean value of the survival probability for the GOE model and for the chaotic spin model, respectively. The entire dynamics is depicted, from the moment the system is quenched out of equilibrium to the moment a new equilibrium is reached, which happens when $\left<P_S(t)\right>$ only fluctuates around a finite asymptotic value. This saturation point corresponds to
\be
\left< \overline{P_S} \right> = \left<  \lim_{t\rightarrow\infty} P_S(t) \right> = \left<\sum_\alpha\left|c_\alpha^0\right|^4\right>.
\ee

The analytical expression for the evolution of the survival probability was obtained for large Hamiltonian matrices in~\cite{Torres2018,SantosTorres2017AIP,Schiulaz2019}. For both the GOE and the realistic model, it is given by
\begin{equation}
\left<P_S(t)\right>=\frac{1- \left< \overline{P_S} \right>}{D-1}\left[D b_1^2( \Gamma t)-b_2\left(\frac{\Gamma t}{\mu D}\right)\right]+ \left< \overline{P_S} \right>.
\label{Eq:PsGOE}
\end{equation}
The first term in the equation above is determined by the shape and energy bounds of the LDOS~\cite{Tavora2016,Tavora2017,Torres2014PRA,Torres2014NJP}, which depend on the model and initial state. For GOE matrices, the shape is semicircular, up to corrections that are subdominant in $1/D$, which gives~\cite{Torres2014PRA,Torres2014NJP}
\be
b^2_1(\Gamma t)=\frac{\mathcal{J}^2_1(2 \Gamma t)}{\Gamma^2 t^2},
\ee
where ${\cal J}_1$ is the Bessel function of the first kind. For realistic chaotic many-body systems, the LDOS is Gaussian~\cite{ZelevinskyRep1996,Flambaum2001a,Kota2001PRE,Torres2014PRA,Torres2014NJP} and $b_1$ involves an early Gaussian decay, $e^{-\Gamma^2 t^2}$, and a later power-law behavior $\propto t^{-2}$, as given in~\cite{Torres2018,Schiulaz2019}. 

The $b_1$ function controls the initial decay of the survival probability.  For $\Gamma t\ll  1$, it leads to the universal $1 - \Gamma^2 t^2$ behavior, where $1/\Gamma$ is the characteristic time for the depletion of the initial state. Later, oscillations emerge that decay as a power law~\cite{Tavora2016,Tavora2017}. The power-law behavior continues until the minimal value of $\left<P_S(t)\right>$ is reached at $t_{\text{Th}}$. This time is referred to as Thouless time and it marks the point of the complete spread of the initial state in the many-body Hilbert space, as explained in Ref.~\cite{Schiulaz2019}.

Beyond $t_{\text{Th}}$, the dynamics become universal and determined by the second term in Eq.~(\ref{Eq:PsGOE}), which is the two-level form factor,
\be
b_2(t) =\begin{cases}
1-2t + t \ln(2 t+1) & t\leq1 \\
t \ln \left( \dfrac{2 t+1}{2 t -1} \right) -1 & t>1
\end{cases}.
\label{eq:b2}
\ee 
This function is responsible for the dip below $\left< \overline{P_S} \right>$, which is known as correlation hole and exists only when the eigenvalues are correlated~\cite{Leviandier1986,Wilkie1991,Alhassid1992,Torres2017Philo}. 
Since the GOE and the realistic chaotic model have similar level statistics, the same equation for $b_2(t)$ is used for both cases. In Eq.~(\ref{Eq:PsGOE}), $\mu=2$ for the GOE model and $\mu = \sqrt{2 \pi}$ for the spin model~\cite{Schiulaz2019}. The $b_2$ function initially grows linearly and later shows a power-law behavior up to saturation, which happens at the relaxation time $t_{\text{R}}$.

We therefore have four regions in time that exhibit different behaviors, as indicated in Fig.~\ref{fig:PsGOE}~(a) and Fig.~\ref{fig:PsGOE}~(b):
\begin{enumerate}
	\item The short time region, for $t \ll 1/\Gamma$.
	\item The power-law decay, happening for $1/\Gamma< t <t_{\text{Th}}$. 
	\item The interval for the correlation hole, $t_{\text{Th}}< t <t_{\text{R}}$. The time $t_{\text{Th}}$ to reach the minimum of the hole is a constant for the GOE model, but grows exponentially with system size for the spin model~\cite{Schiulaz2019}. This exceedingly long time is a consequence of the spatial locality of the initial state and of couplings of the realistic model.
	\item The saturation region, for $t> t_{\text{R}}$. The relaxation time (or Heisenberg time) is the largest time scale of the system and is given by the inverse of the mean level spacing~\cite{Schiulaz2019}. 
\end{enumerate}

Four distinct behaviors, at the same time scales identified for the survival probability, appear also for the spin autocorrelation function, as seen in Fig.~\ref{fig:PsGOE} (e). Similarly to what one finds for $\langle P_S(t) \rangle$, the correlation hole is evident for $\langle I(t) \rangle$ as well. Note, however, that we know analytically that the ratio between the saturation point and the minimum value of $\left<P_S(t)\right>$ remains constant as the system size increases, but this analysis has not been done for $\langle I(t) \rangle$.

For the quantities that are local in time, we observe two different behaviors before the Thouless time, one for $t<1/\Gamma$ and another one for $1/\Gamma< t <t_{\text{Th}}$, as evident in the plots for the inverse participation ratio [Fig.~\ref{fig:PsGOE}~(c)], second-order R\'enyi entropy [Fig.~\ref{fig:PsGOE}~(d)], and connected spin-spin correlation function [Fig.~\ref{fig:PsGOE}~(f)], and also for the Shannon entropy (not shown). However, beyond $t_{\text{Th}}$, the effects of the correlation hole are minor for these quantities, and we basically see only fluctuations around their infinite-time averages. We can then say that the dynamics of $\langle \text{IPR}(t) \rangle $, $\langle S(t) \rangle $, $\langle Sh(t) \rangle $, and $\langle C(t) \rangle $ saturate before the Thouless time.

\section{Survival probability}
\label{Sec:SP}

Despite being a central quantity in the analysis of systems out of equilibrium, not much is known about the self-averaging properties of the survival probability. Some of the existing works have focused on the two-level form factor, which corresponds to the long-time part of the survival probability. They include numerical studies about the spectral correlations of the hydrogen atom in a magnetic field~\cite{Eckhardt1995} and theoretical arguments~\cite{Prange1997}, that both indicate the lack of self-averaging of the two-level form factor.

Here, we provide an analytical expression for ${\cal R}_{P_S}^\textrm{GOE}(t)$ at all times, and show that the survival probability is not self-averaging at any time scale. We confirm numerically that this picture holds for physical chaotic models as well. We start the discussion below with estimates for ${\cal R}_{P_S} (t)$ for both models at short and long times, and then proceed with the presentation of the analytical result for ${\cal R}_{P_S}^\textrm{GOE}(t)$ and numerical results for both models.

\subsection{Short times}

For short times, $t \ll 1/\Gamma $, one can expand the survival probability as
\be
P_S(t)=1-\Gamma_0^2 t^2+{\cal O}(t^4).
\label{SPshortT}
\ee
From this expansion, one finds that the relative variance is given by
\be
{\cal R}_{P_S}=\sigma^2_{\Gamma^2}t^4+{\cal O}(t^6),
\label{eq:sigmaSPshort}
\ee
where $\sigma^2_{\Gamma^2}=\left<\Gamma_0^4\right>-\left<\Gamma_0^2\right>^2$. For the GOE and the realistic chaotic model, one has 
\ba
&& (\sigma^{2}_{\Gamma^2})^{\text{GOE}}=2 D , \nonumber \\ 
&& (\sigma^{2}_{\Gamma^2})^{\text{spin}} = \frac{J^4 L^2 (L-2)}{64 (L-1)^2}  \sim \frac{J^4 }{64 }L ,
\label{eq:sigmaShort}
\ea
so the relative variance grows linearly with matrix and system size, respectively. The survival probability is non-self-averaging for both the GOE and the spin model.

\subsection{Long times}

Strong evidence for the absence of self-averaging of the survival probability at long times was already hinted at by studies of temporal fluctuations. In chaotic systems after saturation, the dispersion of the temporal fluctuations of $P_{S}(t)$ is proportional to the value of the infinite-time average, $\overline{P_{S}}$ \cite{Torres2014NJP,TorresKollmar2015}, so the relative variance remains constant as $L$ increases. The same result is obtained also for the relative variance of $P_{S}(t)$ over the ensemble of realizations, because for chaotic systems and time intervals beyond $t_\text{R}$, temporal averages become comparable to ensemble averages.

The survival probability in Eq.~(\ref{eq:PS}) can also be written as
\be
P_S(t)=\sum_{\alpha \neq \beta} |c_\alpha^{(0)}|^2 |c_\beta^{(0)}|^2e^{-i(E_\alpha-E_\beta)t}  + \sum_{\alpha} | c_\alpha^{(0)} |^4. 
\label{Eq:extra}
\ee
For times $t>t_{\text{R}}$, the ensembles averages cancel out the first term in Eq.~(\ref{Eq:extra}), so $\left< P_S (t>t_{\text{R}}) \right> \sim  \langle \overline{P_S} \rangle =  \langle \sum_{\alpha} | c_\alpha^{(0)} |^4  \rangle $. The eigenvectors of GOE random matrices are statistically equivalent to normalized Gaussian random vectors~\cite{footOrtho}, which gives $\langle \overline{P_S} \rangle \sim 3/D$. For physical models, we also have $\langle \overline{P_S} \rangle \propto 1/D$ for eigenstates away from the borders of the spectrum. 

To obtain the variance $\sigma_{P_S}^2(t)$, in addition to $\langle P_S(t)\rangle ^2$, one needs the ensemble average of the square of the survival probability as well,
\be
\left<P_S^2(t)\right> \!=\! \left<\sum_{\alpha,\beta,\gamma,\delta} \!\!\! e^{-i(E_\alpha-E_\beta+E_\gamma-E_\delta)t}\! \left|c_\alpha^0\right|^2 \! \left|c_\beta^0\right|^2 \! \left|c_\gamma^0\right|^2 \! \left|c_\delta^0\right|^2 \! \right>.
\label{eq:PS2}
\ee
At large times, only the terms with $\alpha=\beta$ and $\gamma=\delta$, or $\alpha=\delta$ and $\gamma=\beta$ matter, because the phase factors average to zero.  Since $2 \langle \sum_{\alpha} | c_\alpha^{(0)} |^4   \sum_{\beta} | c_\beta^{(0)} |^4  \rangle \sim 2 \langle \sum_{\alpha} | c_\alpha^{(0)} |^4  \rangle \langle \sum_{\beta} | c_\beta^{(0)} |^4  \rangle $, one finds that $\sigma^2_{P_S} \propto \langle \overline{P_S} \rangle^2$, and as a consequence
\be
{\cal R_{P_S}}  (t>t_{\text{R}})  ={\cal O}(1).
\ee
At long times, the relative variance is therefore independent of system size.

\subsection{Analytical result for all times}

The derivation of the analytical expression for the mean value of the squared survival probability evolved under random matrices is lengthy, and is explained in Appendix~\ref{app:Analyt}. The formula reads
\ba
 \left<P_S^2(t)\right> &=& g_4(t)+ \langle \overline{P_S} \rangle g_3(t)+4 \langle \overline{P_S} \rangle \left<P_S(t)-   \langle \overline{P_S} \rangle    \right> 
\nonumber \\
&+& \langle \overline{P_S} \rangle^2 \left<P_S(2t) -   \langle \overline{P_S} \rangle   \right>+2  \langle \overline{P_S} \rangle^2  .   
\label{eq:PS2analytic}
\ea
In the above, the functions $g_4(t)$ and $g_3(t)$ are related to the Fourier transforms of the four- and three-point spectral correlation functions, respectively. The agreement between Eq.~(\ref{eq:PS2analytic}) and the numerical results is perfect, as can be seen in the figure in Appendix~\ref{app:Analyt}. Combining Eq.~(\ref{Eq:PsGOE}) and Eq.~(\ref{eq:PS2analytic}), one obtains ${\cal R}_{P_S}^\textrm{GOE}(t)$ analytically. 

\begin{figure}[ht]
\includegraphics*[width=0.47\textwidth]{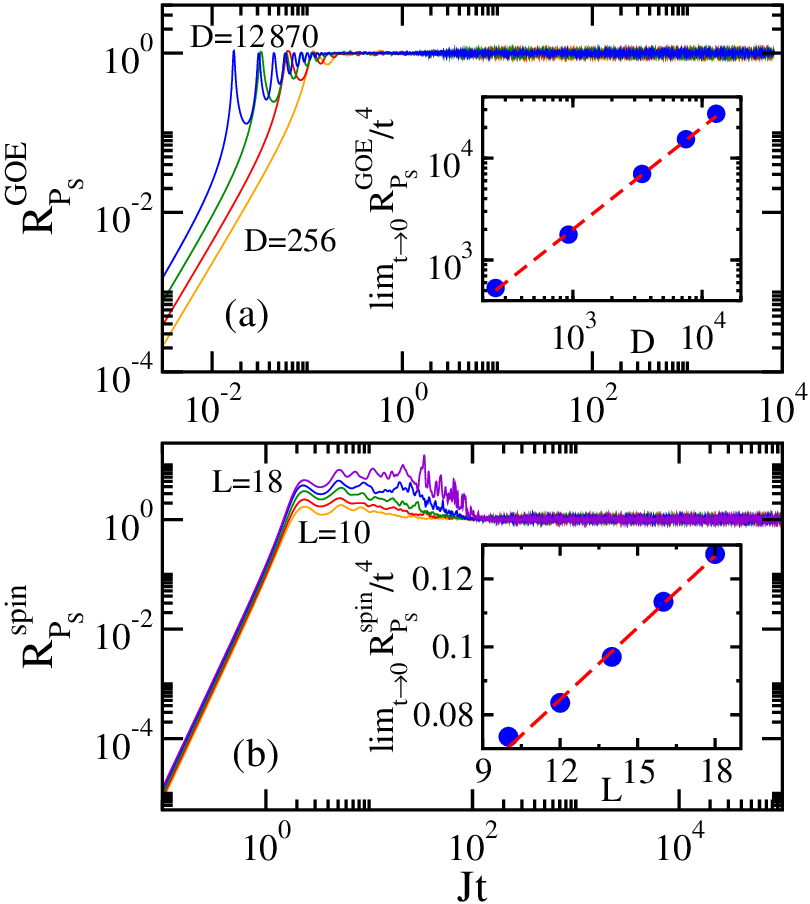}
\caption{Relative variance ${\cal R}_{P_S}(t)$ for the survival probability for the GOE  model (a) and the spin model (b). The system sizes are indicated in the main panels. From the bottom curve to the top curve, the sizes of the matrices are $D=252,924,3\,432,12\,870$ (orange, red, green, blue), and also $D=48\,620$ (purple) for panel (b). The short times coefficients of $\lim_{t\rightarrow0}{\cal R}_{P_S}(t)/t^4$ are plotted for the GOE and spin models in the insets of panels (a) and (b), respectively. In panel (a), the coefficients are shown as a function of $D$, while in panel (b), as a function of $L$.}
\label{fig:SP}
\end{figure}

In Fig.~\ref{fig:SP} (a), we plot the numerical results for ${\cal R}_{P_S}^\textrm{GOE}(t)$ for various matrix sizes.  It is clear that the survival probability is not self-averaging at any time scale. The analytical expression for ${\cal R}_{P_S}^\textrm{GOE}(t)$ agrees very well with the numerics for $t > 1/\Gamma$, while for very short times, where the Fourier transform of the LDOS dominates the dynamics, finite size corrections are important. In this case, one computes ${\cal R}^\textrm{GOE}_{P_S}(t)$ using Eq.~(\ref{eq:sigmaSPshort}). This is shown in the inset of Fig.~\ref{fig:SP} (a), where the dots are the coefficients for $\lim_{t\rightarrow0}{\cal R}_{P_S}^\textrm{GOE}(t)/t^4$ extracted numerically and the dashed line shows the analytical behavior predicted from Eq.~(\ref{eq:sigmaSPshort}) and Eq.~(\ref{eq:sigmaShort}), $\lim_{t\rightarrow0}{\cal R}_{P_S}^\textrm{GOE}(t)/t^4 \propto D$. 

A qualitatively analogous behavior is shown for the spin model in Fig.~\ref{fig:SP} (b). At short times, ${\cal R}^\textrm{spin}_{P_S}(t)$ grows with system size, and at large times, it is  size independent. The inset confirms the prediction by Eq.~(\ref{eq:sigmaSPshort}) and Eq.~(\ref{eq:sigmaShort}), indicating that  $\lim_{t\rightarrow0}{\cal R}_{P_S}^\textrm{spin}(t)/t^4$ is proportional to $L$. This coefficient grows slower for the spin model than for the GOE model, because of the sparseness of the realistic Hamiltonian. 

At intermediate times, where in Fig.~\ref{fig:PsGOE}, $\left< {P_S}(t) \right>$ shows oscillations decaying as a power law, ${\cal R}_{P_S}(t)$ also oscillates for both models, as observed in Fig.~\ref{fig:SP}. In the case of the physical model, the relative variance in this region reaches values above 1. Since the power-law decay of $\left< {P_S}(t) \right>$ is caused by the bounds of the spectrum of finite systems~\cite{Tavora2016,Tavora2017}, at such intermediate times,  the state $\left|\Psi(t)\right>$ acquires weight on eigenstates closer to the edges of the spectrum, which are not described by random matrix theory. The values of ${\cal R}^\textrm{spin}_{P_S}(t)$ above 1 could be a manifestation of correlations between the components of these states. 

Beyond the region of the power-law decay, the relative variance of the survival probability behaves similarly for any time, ${\cal R}_{P_S}(t>t_{\text{Th}}) \sim 1$. This suggests that the correlation hole, which is clearly manifested in the mean value of the survival probability in Fig.~\ref{fig:PsGOE}, does not affect the self-averaging properties of this quantity in any way different from what one sees for $t>t_{\text{R}}$.

\section{Inverse Participation Ratio and R\'enyi Entropies}
\label{Sec:IPR}

The inverse participation ratio and R\'enyi entropies are defined in Eq.~(\ref{eq:IPR}), Eq.~(\ref{eq:Sr}), and Eq.~(\ref{eq:Sh}). While their self-averaging properties are similar for $t>1/\Gamma$, they differ for short times.

\subsection{Inverse participation ratio}
There is a fundamental difference between $P_S(t)$ and $\text{IPR}(t)$: the survival probability measures the distance of the evolved state $\left|\Psi(t)\right>$ from the initial state (indicated as $n=0$), while the inverse participation ratio measures the distance of $\left|\Psi(t)\right>$ from any unperturbed many-body state $\left|n\right>$. The inverse participation ratio can be seen as a generalization of the squared survival probability, where in addition to the term with $n=0$, which gives $P_S^2(t)$ itself, it contains also all other terms with $n\neq 0$. This parallel allows us to intuitively understand the behavior of the relative variance of IPR. At short times, $\left|\Psi(t)\right>$ is still very close to $\left|\Psi(0)\right>$, so the term with $n=0$ dominates the evolution and $\text{IPR}(t)$ behaves analogously to $P_S^2(t)$ \cite{Borgonovi2019b}. This means that self-averaging is absent at short times. At large times, on the other hand, the average over all unperturbed many-body states drastically reduces the fluctuations, and the inverse participation ratio becomes very strongly self-averaging, in the sense that ${\cal R}_{\text{IPR}}\propto D^{-1}$ for both the GOE and the spin model. 

For times $t \ll 1/\Gamma $, $\text{IPR}(t)$ can be expanded as
\be
\text{IPR}(t)=1-2\Gamma_0^2t^2+{\cal O}(t^4),
\label{eq:IPRshortT}
\ee
which is exactly the same expression one has for $P_S^2(t)$ at leading order in $t$. This means that 
\be
{\cal R}_{\text{IPR}}(t) \propto {\cal R}_{P_S}(t) \propto\sigma_{\Gamma^2}^2t^4+{\cal O}(t^6)
\label{Eq:RIPRshort}
\ee
at short times, for any model, so the same kind of non-self-averaging behavior found in Sec.~\ref{Sec:SP} emerges here also.

At large times, we study the ensemble average of 
\ba
\text{IPR}(t)&=&\sum_n\sum_{\alpha,\beta,\gamma,\delta}e^{-i(E_\alpha-E_\beta+E_\gamma-E_\delta)t}\nonumber\\
&\times& c_\alpha^0 c_\alpha^{n*} c_\beta^n c_\beta^{0*} c_\gamma^0 c_\gamma^{n*} c_\delta^n c_\delta^{0*}
\label{eq:IPRfull}
\ea
and  of $\text{IPR}^2(t)$ using the same arguments employed in Sec.~\ref{Sec:SP}, namely that the eigenstates of GOE matrices imply that $\left< c_\alpha \right> \sim 0 $, $\left< |c_\alpha|^2 \right> \sim 1/D $, and $\left< \sum_{\alpha ,\beta} e^{-i(E_\alpha-E_\beta)t} \right> \sim 0 $ unless $\alpha = \beta$. One finds that
\be
{\cal R}_{\text{IPR}}^{\text{GOE}} (t > t_{\text{R}}) \propto\frac{1}{D}.
\label{Eq:RIPRgoe}
\ee
Thus, unlike the survival probability, $\text{IPR}(t)$ is actually self-averaging at large times.

\begin{figure}[h!]
\includegraphics*[width=0.47\textwidth]{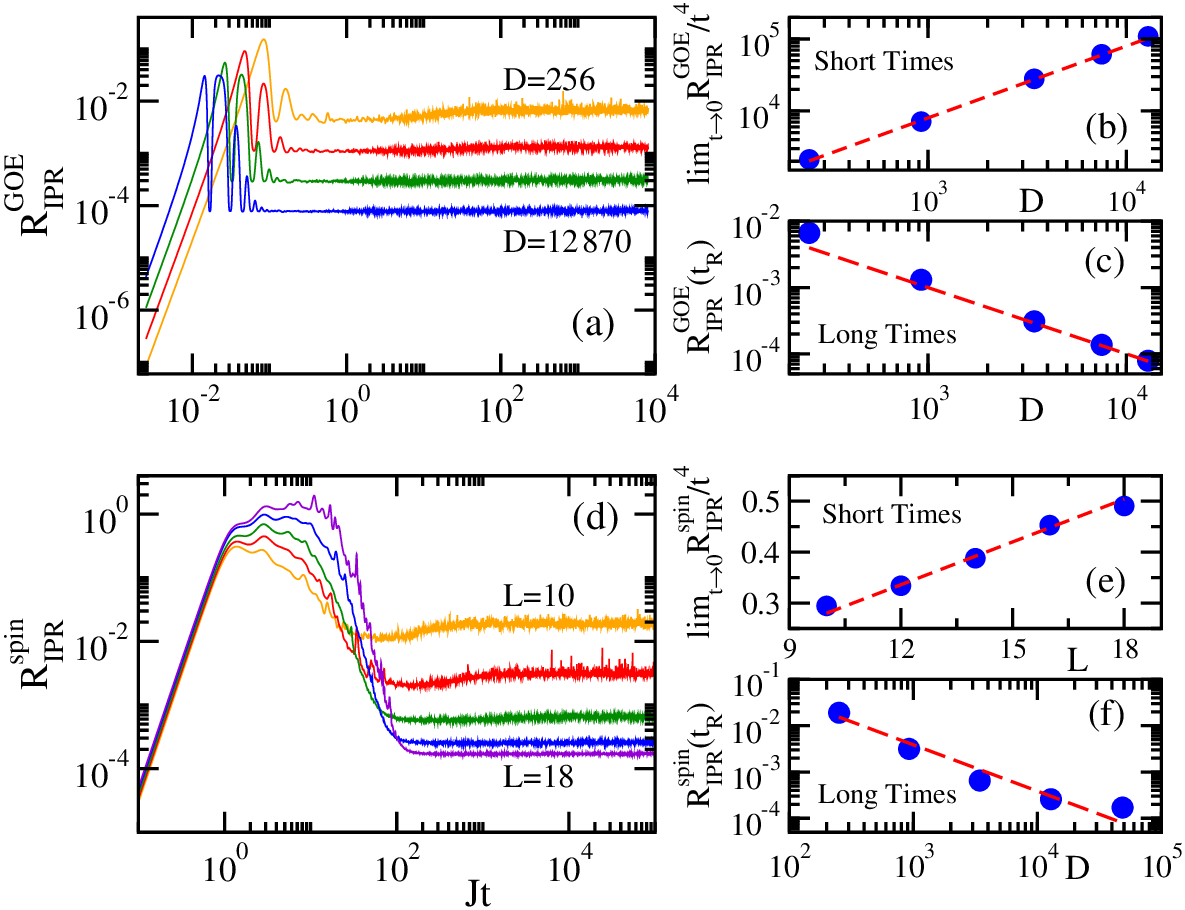}
\caption{Relative variance of the inverse participation ratio for the GOE (a) and spin model (d). The system sizes are indicated in (a) and (d). From the top curve to the bottom curve (at large times), the sizes of the matrices are $D=252,924,3\,432,12\,870$ (orange, red, green, blue) and also $D=48\,620$ (purple) in (d). In (b) and (e): coefficient $\lim_{t\rightarrow0}{\cal R}_{\text{IPR}}(t)/t^4$ as a function of $D$ for the GOE model (b), and as a function of $L$ for the spin model (e). In (c) and (f): values of ${\cal R}_{\text{IPR}} (t \geq t_\text{R})$ at long times as a function of $D$ for the GOE (c) and spin model (f). In (b), (c), (e), and (f), the circles are for numerical data and they agree well with theoretical estimates (dashed lines).}
\label{fig:IPR}
\end{figure}

Not all eigenstates of realistic chaotic models are close to normalized Gaussian random vectors, but they are the majority, so we should expect a similar behavior for the realistic spin model as well.

In Fig.~\ref{fig:IPR}, we plot the numerical data for ${\cal R}_{\text{IPR}}^\textrm{GOE}(t)$ in panel (a), and for ${\cal R}_{\text{IPR}}^\textrm{spin}(t)$ in panel (d). In both cases, the lack of self-averaging at short times and the very strong self-averaging at large times are clearly visible. This is shown quantitatively in the other panels. In Fig.~\ref{fig:IPR} (b) and (e), we plot the short time coefficient $\lim_{t\rightarrow0}{\cal R}_{\text{IPR}}(t)/t^4$ for the GOE and the spin model, respectively. In Fig.~\ref{fig:IPR} (c) and (f), we show the value of ${\cal R}_{\text{IPR}}(t)$ for a long time $t \geq t_{\text{R}}$.  In all four cases, the numerical values (circles) agree very well with our analytical estimates (dashed lines) in Eq.~(\ref{Eq:RIPRshort}) and Eq.~(\ref{Eq:RIPRgoe}), except for a small noticeable deviation for the spin model with $L=18$ in Fig.~\ref{fig:IPR} (f).

At intermediate times, in the region of the power-law decay of $P_S(t)$ and similarly to the behavior of ${\cal R}_{P_S}(t)$, the relative variance of the inverse participation ratio oscillates and reaches its largest values, as seen in Fig.~\ref{fig:IPR}~(a) and Fig.~\ref{fig:IPR}~(d). Beyond this region, the curves are pretty much flat and ${\cal R}_{\text{IPR}}$ is  unaware of the correlation hole.

\subsection{R\'enyi entropies}

Similarly to the behavior of the inverse participation ratio, the second-order R\'enyi entropy is super self-averaging at long times and non-self-averaging at the time scales of the power-law decay of $P_S(t)$ (see figures for ${\cal R}_{S}(t)$ in Appendix~\ref{app:Rs}). The two quantities differ, however, at short times. This happens, because for $t \ll 1/\Gamma $, 
\be
S(t)=2\Gamma_0^2t^2+{\cal O}(t^4),
\ee
so $S(t)\rightarrow0$ for $t\rightarrow0$, while the inverse participation ratio goes to 1 for $t\rightarrow0$ [see Eq.~(\ref{eq:IPRshortT})].  Contrary to ${\cal R}_{\text{IPR}}(t)$, the time dependence of ${\cal R}_S(t)$ cancels out at the lowest order in $t$,
\be
{\cal R}_S(t)=\frac{\sigma^2_{\Gamma^2}}{\left<\Gamma_0^2\right>^2}+{\cal O}(t^2).
\ee
Using Eq.~(\ref{Eq:GammaGOE}), Eq.~(\ref{Eq:GammaSpin}), and Eq.~(\ref{eq:sigmaShort}), one finds that the second-order R\'enyi entropy is self-averaging at short times for both random matrices and physical models,
\be
{\cal R}_S^\textrm{GOE}(t)=\frac{2}{D}+{\cal O}(t^2), \quad {\cal R}_S^\textrm{spin}(t)=\frac{1}{L}+{\cal O}(t^2).
\label{Eq:SatShortTime}
\ee

The difference in the behavior of $\text{IPR(t)}$ and $-\ln [\text{IPR(t)}]$ is somewhat reminiscent to what happens in  the Anderson model, where the transmission amplitude, which scales multiplicatively with the system size, is not self-averaging, while its logarithm, which scales additively, is self-averaging~\cite{MullerARXIV}. The fact that  $S(t)$ is self-averaging at short times makes it more appealing for experiments than $\text{IPR(t)}$. 

From the point of view of self-averaging properties, whether one uses the second-order R\'enyi entropy or the Shannon entropy is indifferent. We verified that both ${\cal R}_S(t)$ and ${\cal R}_{Sh}(t)$ exhibit equivalent behaviors.

\section{Experimental local quantities}
\label{Sec:Imb}

The two experimental quantities considered, the spin autocorrelation function and the connected spin-spin correlation function, are local in space. Since for random matrices, the notion of locality is meaningless, we study these quantities for the spin model only. 

The main difference between the two observables is that the connected spin-spin correlation function is also local in time, while the spin autocorrelation function is not. In this sense, the spin autocorrelation function is the spatially local counterpart of the survival probability, since both are measured with respect to the state of the system at $t=0$, and the connected correlation function is the spatially local counterpart of the inverse participation ratio, both involving only the state at $t$ and both dealing also with averages over all unperturbed many-body states.

The differences between these quantities are reflected in their self-averaging properties. At short times, spatial locality ensures that both observables are self-averaging. Their dynamics involve only a finite number of spins, and the spatial averages ensure that the relative variances are reduced as the system size increases. At long times, on the other hand, the spin autocorrelation function, just as the survival probability, is not self-averaging, while the connected correlation function, similarly to the inverse participation ratio, is.

\subsection{Spin autocorrelation function}

At short times, $t \ll 1/\Gamma $, one can expand the spin autocorrelation function in the following way,
\be
I(t) \!=\!1-\Gamma_0^2t^2+\frac{4t^2}{L}\sum_{k=1}^LS_k^{00} \sum_{n\neq0} \left| \langle n|H|\Psi(0)\rangle \right|^2 S_k^{nn}+{\cal O}(t^4),
\label{eq:Ishort}
\ee
where $S_k^{nn}=\left<n\right|S_k^z\left|n\right>$. The third term on the right-hand side of Eq.~(\ref{eq:Ishort}) is zero, unless $|n\rangle$ is directly coupled with $|\Psi(0)\rangle $. These states $|n\rangle$ differ from $|\Psi(0)\rangle $ by two neighboring sites only. Therefore, using the definition of $\Gamma_0^2$ in Eq.~(\ref{eq:Gamma}), one finds that 
\[
\sum_{k=1}^LS_k^{00} \sum_{n\neq0} \left| \langle n|H|\Psi(0)\rangle \right|^2S_k^{nn} = \frac{L-4}{4} \Gamma_0^2 ,
\]
which gives 
\be
I(t)=1-\frac{4  \Gamma_0^2  t^2}{L}+{\cal O}(t^4).
\ee
From this expansion, we obtain the relative variance, 
\be
{\cal R}_I(t) = \frac{16 \sigma_{\Gamma^2}^2 t^4}{L^2}+{\cal O}(t^6) \propto \frac{J^4 t^4}{L},
\ee
where we used that $\sigma_{\Gamma^2}^2 \propto L$ [see Eq.~(\ref{eq:sigmaShort})]. 

The estimates above provide us with two important results about the spin autocorrelation function at short times: (i) Its mean value $\left<I(t)\right>$ is independent of system size, since according to Eq.~(\ref{Eq:GammaSpin}), $\left< \Gamma_0^2 \right> \propto L$; and (ii) the quantity is strongly self-averaging, since the relative variance decays as $1/L$. Both these features are a consequence of the local structure of the observables and of the Hamiltonian, and not of any peculiar property of $I(t)$. As a consequence, we can claim that the local quantities studied here, or any other involving only sums of local operators and evolving under local Hamiltonians, are self-averaging for short times.

\begin{figure}[h!]
\hspace{-0.4 cm}
\includegraphics[width=0.49\textwidth]{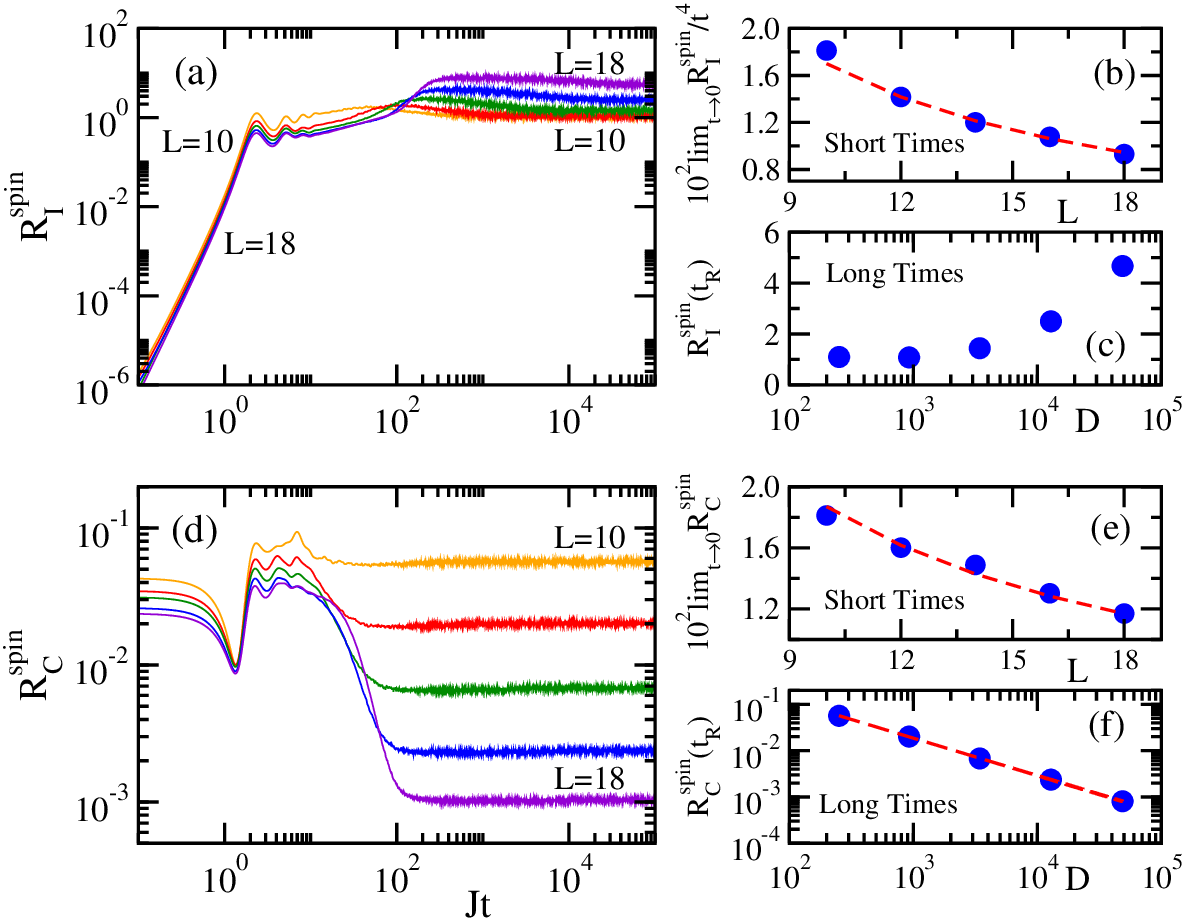}
\caption{Relative variance of the spin autocorrelation function (a) and connected spin-spin correlation function (d) for the spin model. The system sizes are indicated in (a) and (d): From the top curve to the bottom curve (at short times), the sizes of the matrices are $D=252,924,3\,432,12\,870,48\,620$ (orange, red, green, blue, purple). This changes for the spin autocorrelation function at long time. Panels (b) and (e) depict, respectively,  the short time coefficient $10^2 \times \lim_{t\rightarrow0} {\cal R}_{I}(t)/t^4$ and $10^2 \times \lim_{t\rightarrow0} {\cal R}_{C}(t)$ as a function of system size $L$. The numerical data (circles) are compared with a fitting curve $\propto 1/L$ (dashed line). Panels (c) and (f) show numerical data (circles) for ${\cal R}_{I}(t)$ and ${\cal R}_{C}(t)$ for a time $t>t_{\text{R}}$  as a function of the dimension $D$; in (f): fitting curve $\propto 1/D^{0.8}$ (dashed line).}
\label{fig:I}
\end{figure}

The relative variance ${\cal R}_I(t)$ is plotted  in Fig.~\ref{fig:I} (a). At short times, we observe the expected power-law behavior $\propto t^4$, with a coefficient $\lim_{t\rightarrow0} {\cal R}_I(t)/t^4\propto 1/L$, as shown in Fig.~\ref{fig:I} (b). The self-averaging behavior persists up to times currently reachable experimentally, so self-averaging can be safely assumed in real experiments. 

At large times, however, an inversion happens, and ${\cal R}_I(t)$ starts growing with system size, as seen in Fig.~\ref{fig:I}~(a) and Fig.~\ref{fig:I}~(c). This happens at times of the order of the Thouless time, when $\langle I(t) \rangle$ enters the correlation hole [see Fig.~\ref{fig:PsGOE}~(e)], and the dynamics crossover from a model-dependent regime at short times, to a universal regime at long times~\cite{Schiulaz2019}.  After this point, self-averaging is lost, and in an even stronger sense than for the survival probability, for which ${\cal R}_{P_S}(t)$ for $t > t_{\text{Th}}$ is system size independent. At these very long time scales, the spin autocorrelation function presents positive and negative values very close to zero, which results in large fluctuations.

\subsection{Connected spin-spin correlation function}

Using a short-time expansion for the connected spin-spin correlation function, one finds that, 
\ba
C(t) & \!=\! &  \frac{4 t^2}{L} \sum_{k=1}^L  \sum_{n \neq 0}
 \left( S_k^{00} S_{k+1}^{00}  - 2 S_k^{00} S_{k+1}^{nn} + S_k^{nn} S_{k+1}^{nn} \right) \times
    \nonumber \\
&&   \left| \langle n|H|\Psi(0)\rangle \right|^2 +{\cal O}(t^4)  = -\frac{2 \Gamma_0^2 t^2}{L} +{\cal O}(t^4).
\ea
This implies that, as a consequence of space locality and since $\langle \Gamma_0^2 \rangle \propto L$, the mean value of this quantity is system size independent, just like the spin autocorrelation function, and it is also self-averaging, 
\be
{\cal R}_C(t) = \frac{ \sigma_{\Gamma^2}^2 }{\langle \Gamma_0^2 \rangle^2} +{\cal O}(t^6)
\sim \frac{1}{L}.
\ee

The main difference between the connected spin-spin correlation function and the spin autocorrelation at short times is that $C(0)=0$, while $I(0)=1$. Therefore, the relative variance ${\cal R}_C(t)$ tends to a finite value as $t\rightarrow0$, while ${\cal R}_I(t)\rightarrow0$. In Fig.~\ref{fig:I}  (d), we plot ${\cal R}_C(t)$ for the spin model. At short times, the relative variance indeed saturates to a finite value. Figure~\ref{fig:I}~(e) confirms that this value decays as $1/L$.

At large times, ${\cal R}_C (t)$ saturates to an asymptotic value that decreases exponentially with $L$, as shown in Fig.~\ref{fig:I}~(f). This is similar to what happens for the inverse participation ratio, and in contrast to the behavior of the spin autocorrelation function. We conjecture that this very strong self-averaging behavior at long times is associated to the fact that no memory of the initial state is encoded in the connected spin-spin correlation function.

\section{Impact of Lack of Self-Averaging}
\label{Sec:Lack}

Lack of self-averaging, and therefore not being able to decrease the number of random realizations (or samples) as the system size $L$ increases, can make numerical and experimental analysis quite challenging. It is also very dangerous to assume self-averaging, as one often does when studying chaotic systems, and then decrease the number of realizations as $L$ increases for a quantity that is in fact non-self-averaging. The purpose of this section is to provide an example of this case.

\begin{figure}[h!]
\hspace{-0.4 cm}
\includegraphics[width=0.4\textwidth]{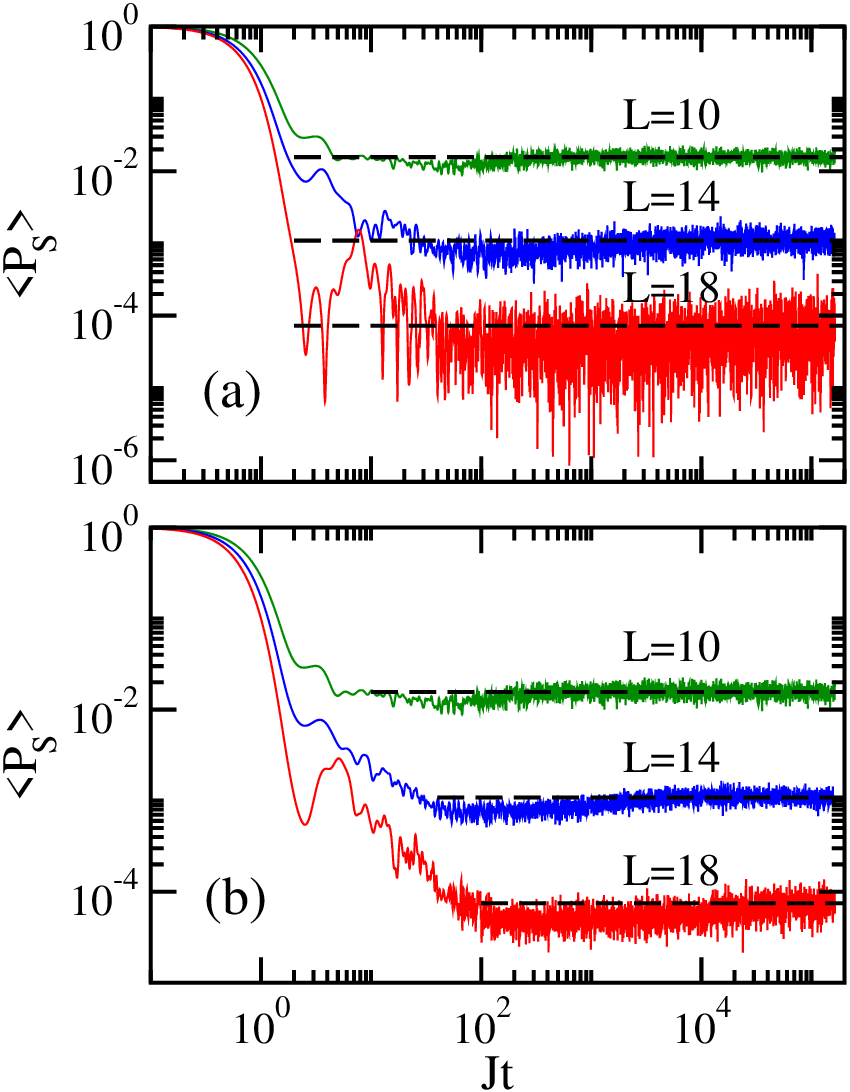}
\caption{Survival probability averaged over a number of random realizations that decreases (a) or remains constant (b) as the system size increases. Two initial states in the middle of the spectrum are considered in all cases. The system sizes are indicated in the figure. In panel (a): $\langle P_S (t)\rangle$ for $L=10$ is averaged over 32 random realizations, for $L=14$ over 8 random realizations, and for $L=18$ over 2 random realizations. In panel (b): all system sizes count with 32 random realizations. The horizontal dashed line marks the infinite time average $\langle \overline{P_S} \rangle $.}
\label{fig:Impact}
\end{figure}

In Fig.~\ref{fig:Impact}, we show the survival probability for the chaotic spin model (\ref{eq:Hspin}) averaged over a number of random realizations. This number decreases as $L$ increases in Fig.~\ref{fig:Impact}~(a), but remains constant as $L$ increases in Fig.~\ref{fig:Impact}~(b). Main features of the evolution of the survival probability are missed in Fig.~\ref{fig:Impact}~(a). There is a hint for the correlation hole for $L=10$, which is entirely lost for $L=18$. There is an indication of a power-law decay before the correlation hole for $L=14$, but this is also lost for $L=18$. As a result, motivated by the curve for the largest system size, one could assume that what one sees for the smaller sizes are just finite size effects, and that after the Gaussian decay, the dynamics simply saturates. Worse still, since $\Gamma \propto \sqrt{L}$ and the Gaussian decay $e^{-\Gamma^2 t^2}$ gets faster as $L$ increases, one could be led to conclude that the relaxation time decreases as the chain grows, while in reality the relaxation time for $\langle P_S (t) \rangle$ increases exponentially with $L$ (see Sec.~\ref{Sec:mean} and Ref.~\cite{Schiulaz2019}).

In Fig.~\ref{fig:Impact}~(b), both the power-law decay and the correlation hole are visible for $L=14$ and $L=18$ confirming that these features are indeed present. The results would be even more convincing if instead of keeping the number of realizations fixed, we would increase it.

A critical ongoing debate in the current literature about nonequilibrium quantum dynamics is whether existing conclusions may be incomplete or even incorrect due to finite size effects. Our results show that self-averaging is another important element that needs to be taken into account. For a quantity that is non-self-averaging, one can reach wrong conclusions if one assumes self-averaging and decreases the number of realizations.

\section{Conclusions}
\label{Sec:Conc}

This work analyzes the self-averaging behavior of many-body quantum systems out of equilibrium. The focus is on the chaotic regime, where self-averaging is often taken for granted. By examining different non-local and local quantities in space, we bring forward a rich variety of behaviors and deduce that self-averaging is not an intrinsic consequence of quantum chaos, but depends strongly on the quantity and on the time scale. 

On the bright side, the local quantities studied here and measured in experiments with cold atoms and ion traps, namely the spin autocorrelation function and the connected spin-spin correlation function, are self-averaging for the times that are currently experimentally reachable. The same arguments that we employ for these quantities can be extended to any observable comprising only sums of spatially local operators and evolving under local Hamiltonians, so they should also be self-averaging at short times, a result that is reassuring for experimentalists. 

Numerical studies, on the other hand, where long times and an array of quantities are available, should be cautious. Autocorrelation functions, such as the survival probability and the spin autocorrelation function, are not self-averaging at long times, so one needs large statistics even when pushing towards very large system sizes. In fact, as we showed analytically for full random matrices, the survival probability is not self-averaging at any time scale. Extra care should therefore be taken when dealing with this quantity, which has a central role in studies of nonequilibrium quantum dynamics.

The time evolution of the fluctuations of observables is still uncharted territory. There are multiple interesting directions that the study initiated here could take, from the analysis of non-chaotic models, such as those approaching many-body localization, to time-dependent Hamiltonians and open systems. Our results lay the groundwork for such future analysis.

\begin{acknowledgments}
M.S. and L.F.S. were supported by the NSF Grant No.~DMR-1603418. E.J.T.-H. acknowledges funding from VIEP-BUAP (Grant Nos. MEBJ-EXC19-G, LUAGEXC19-G), Mexico. He is also grateful to LNS-BUAP for allowing use of their supercomputing facility. F.P.B. thanks the Consejer\'ia de Conocimiento, Investigaci\'on y Universidad, Junta de Andaluc\'ia and European Regional Development Fund (ERDF), ref. SOMM17/6105/UGR. Additional computer resources supporting this work were provided by the Universidad de Huelva CEAFMC High Performance Computer  located in the Campus Universitario el Carmen and funded by FEDER/MINECO project UNHU-15CE-2848. LFS is supported by the NSF Grant No.~DMR-1936006. Part of this work was performed at the Aspen Center for Physics, which is supported by the NSF grant PHY-1607611.
\end{acknowledgments}

\appendix

\section{Analytical expression for $\pmb{ \left<P_S^2(t)\right>} $} 
\label{app:Analyt}

We provide here the analytical derivation for ${\cal R}_{P_S}^\textrm{GOE}(t)$. The analytical expression for $\left<P_S(t)\right>$ is given in Eq.~(\ref{Eq:PsGOE}) and was already obtained in~\cite{Torres2018,SantosTorres2017AIP}. It remains to show how we obtain Eq.~(\ref{eq:PS2analytic}) for $\left<P_S^2(t)\right>$, which is a much more involved derivation.

\begin{figure}[h!]
\includegraphics*[width=0.45\textwidth]{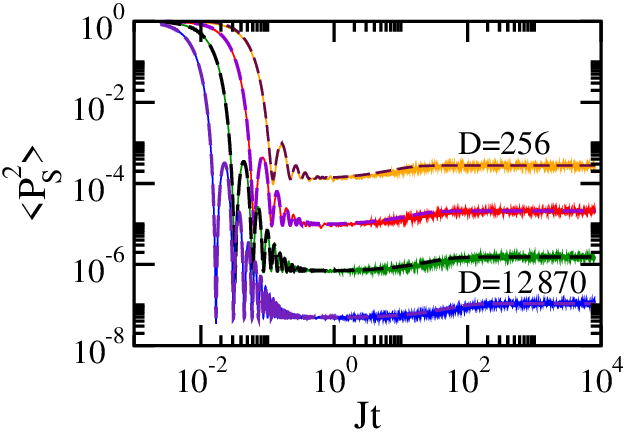}
\caption{Evolution of the squared survival probability for the GOE  model. The system sizes are indicated in the figure. From the top curve to the bottom curve, the sizes of the matrices are $D=252,924,3\,432,12\,870$ (orange, red, green, blue). Numerical results (solid lines) and the analytical expression (dashed lines) from Eq.~(\ref{eq:PS2analytic}) are presented. Averages over $10^4$ data.}
\label{fig:Ps2}
\end{figure}

Before presenting the steps of the derivation, we compare in Fig.~\ref{fig:Ps2}  the expression from Eq.~(\ref{eq:PS2analytic}) (dashed lines) with the numerical results (solid lines) for GOE matrices of different sizes. The agreement is indeed perfect.

To obtain the relative variance of the survival probability, we need to compute
\be
\left<P_S^2(t)\right> = \Xi_1 + \Xi_2 + \Xi_3 + \Xi_4,
\label{eq:PS2expanded}
\ee
where
\ba
\Xi_1 &=& \left<\sum_{\alpha\neq\gamma\neq\beta\neq\delta} \!\!\!\!\!\!\!
e^{-i(E_\alpha-E_\beta+E_\gamma-E_\delta)t}\left|c^{(0)}_\alpha\right|^2\left|c^{(0)}_\beta\right|^2\left|c^{(0)}_\gamma\right|^2\left|c^{(0)}_\delta\right|^2\right> ,
\nonumber \\
\Xi_2 &=& \left<\sum_{\alpha\neq\beta\neq\gamma}e^{-i(2E_\alpha-E_\beta-E_\gamma)t}\left|c^{(0)}_\alpha\right|^2\left|c^{(0)}_\beta\right|^4\left|c^{(0)}_\gamma\right|^2\right> \nonumber\\
&+&\left<\sum_{\alpha\neq\beta\neq\gamma}e^{-i(E_\alpha-2E_\beta+E_\gamma)t}\left|c^{(0)}_\alpha\right|^4\left|c^{(0)}_\beta\right|^2\left|c^{(0)}_\gamma\right|^2\right> ,\nonumber \\
\Xi_3 &=& \left<\sum_{\alpha\neq\beta}e^{-2i(E_\alpha-E_\beta)t}\left|c^{(0)}_\alpha\right|^4\left|c^{(0)}_\beta\right|^4\right> \nonumber \\
&+& 4\left<\sum_{\alpha\neq\beta}e^{-i(E_\alpha-E_\beta)t}\left|c^{(0)}_\alpha\right|^2\left|c^{(0)}_\beta\right|^2\sum_\gamma\left|c^{(0)}_\gamma\right|^4\right>  , \nonumber \\
\Xi_4 &=& 2\left<\sum_{\alpha,\beta}\left|c^{(0)}_\alpha\right|^4\left|c^{(0)}_\beta\right|^4\right> .  \nonumber
\ea
Equation~(\ref{eq:PS2expanded}) is obtained by splitting the sum in Eq.~(\ref{eq:PS2}) into all possible combination of equal indexes $\alpha,\beta,\gamma,\delta$. For example, $\Xi_1$ contains the terms with all indexes different, $\Xi_2$ the terms with either $\alpha=\gamma$ or $\beta=\delta$, and so on. We now compute each one of these terms, starting from $\Xi_4$ and moving upwards.

\subsection{Term $\pmb{\Xi_4}$}

The fourth term in Eq.~(\ref{eq:PS2expanded}) can be computed straightforwardly, applying the results of Refs.~\cite{Torres2018,SantosTorres2017AIP}. Indeed, up to subleading corrections in $D^{-1}$, one finds
\be
2\left<\sum_{\alpha,\beta}\left|c^{(0)}_\alpha\right|^4\left|c^{(0)}_\beta\right|^4\right>=2\left<\overline{P_S}\right>^2+O(D^{-3}).
\ee
Note that, at large times, this is the only term that does not vanish. This is because it is the only term which does not contain fluctuating phases in time. We can therefore compute the asymptotic value of the relative variance already,
\ba
\lim_{t\rightarrow\infty}\mathcal{R}_{P_S}^\textrm{GOE}(t)&=&\frac{2\left<\overline{P_S}\right>^2-\left<\overline{P_S}\right>^2}{\left<\overline{P_S}\right>^2}+O(D^{-1})\nonumber\\
&=&1+O(D^{-1}).
\ea
This means that, as explained in Sec.~\ref{Sec:SP}, the relative variance is system size independent at large times, and self-averaging is not present.

\subsection{Term $\pmb{\Xi_3}$}
 Let us first compute the following term of $\Xi_3$,
 \be
 4\left<\sum_{\alpha\neq\beta}e^{-i(E_\alpha-E_\beta)t}\left|c^{(0)}_\alpha\right|^2\left|c^{(0)}_\beta\right|^2\sum_\gamma\left|c^{(0)}_\gamma\right|^4\right>.
 \ee
 In order to do this, let us define the time dependent part of the survival probability
 \be
 \widetilde{P}_S(t)=P_S(t)-\overline{P_S}=\sum_{\alpha\neq\beta}e^{-i(E_\alpha-E_\beta)t}\left|c^{(0)}_\alpha\right|^2\left|c^{(0)}_\beta\right|^2,
 \ee
 which tends to zero as $t\rightarrow\infty$. From Eq.~(\ref{Eq:PsGOE}), one gets for its average value
 \be
\left<\widetilde{P}_S(t)\right>=\frac{1-\left<\overline{P_S}\right>}{D-1}\left[D b_1^2( \Gamma t)-b_2\left(\frac{\Gamma t}{2D}\right)\right].
\ee
As a consequence, one finds, up to subleading corrections in $D^{-1}$, that
\ba
 &&4 \left<\sum_{\alpha\neq\beta}e^{-i(E_\alpha-E_\beta)t}\left|c^{(0)}_\alpha\right|^2\left|c^{(0)}_\beta\right|^2\sum_\gamma\left|c^{(0)}_\gamma\right|^4\right>\nonumber\\
 &=&4\left<\overline{P_S}\right>\left<\widetilde{P}_S(t)\right>+O(D^{-2}).
\ea

We now compute the other term of $\Xi_3$,
\be
\left<\sum_{\alpha\neq\beta}e^{-2i(E_\alpha-E_\beta)t}\left|c^{(0)}_\alpha\right|^4\left|c^{(0)}_\beta\right|^4\right>.
\ee
For this, one needs to recall that for GOE matrices, eigenvalues and eigenvectors are statistically independent~\cite{MehtaBook}. This means that the averages over the components and the eigenenergies factorize. This fact will be used multiple times in this derivation. For this particular term, it implies that
\ba
&&\left<\sum_{\alpha\neq\beta}e^{-2i(E_\alpha-E_\beta)t}\left|c^{(0)}_\alpha\right|^4\left|c^{(0)}_\beta\right|^4\right> \nonumber \\
&=&\left<\sum_{\alpha\neq\beta}\left|c^{(0)}_\alpha\right|^4\left|c^{(0)}_\beta\right|^4\right>\left<e^{-2i(E_\alpha-E_\beta)t}\right>\nonumber\\
&=&\left<\widetilde{P}_S(2t)\right>\frac{\left<\overline{P_S}\right>^2+O(D^{-3})}{1-\left<\overline{P_S}\right>}\\
&\sim&\left<\widetilde{P}_S(2t)\right>\left(\left<\overline{P_S}\right>^2+O(D^{-2})\right).\nonumber
\ea

\subsection{Term $\pmb{\Xi_2}$}

We now need to compute $\Xi_2$,
\ba
\left<\sum_{\alpha\neq\beta\neq\gamma}e^{-i(2E_\alpha-E_\beta-E_\gamma)t}\left|c^{(0)}_\alpha\right|^2\left|c^{(0)}_\beta\right|^4\left|c^{(0)}_\gamma\right|^2\right>&+&\nonumber\\
\left<\sum_{\alpha\neq\beta\neq\gamma}e^{-i(E_\alpha-2E_\beta+E_\gamma)t}\left|c^{(0)}_\alpha\right|^4\left|c^{(0)}_\beta\right|^2\left|c^{(0)}_\gamma\right|^2\right>.
\ea
Using again the statistical independence of eigenvalues and eigenvectors, we get for the components of the initial state,
\be
\left<\sum_{\alpha\neq\beta\neq\gamma}\left|c^{(0)}_\alpha\right|^2\left|c^{(0)}_\beta\right|^4\left|c^{(0)}_\gamma\right|^2\right>=\left<\overline{P_S}\right>+O(D^{-2}).
\ee

It remains to obtain the following function,
\be
g_3(t)=\left<e^{-i(2E_\alpha-E_\beta-E_\gamma)t}+e^{-i(E_\alpha-2E_\beta+E_\gamma)t}\right>.
\ee
Following Refs.~\cite{Torres2018,SantosTorres2017AIP,Cotler2017GUE}, we write this average as
\ba
g_3(t)&=&\frac{(D-3)!}{D!}\int dE e^{-iEt}\int dE_\alpha dE_\beta dE_\gamma R_3(E_\alpha,E_\beta,E_\gamma)\nonumber\\
&\times&\left[\delta(E-(2E_\alpha-E_\beta-E_\gamma))+\delta(E-(E_\alpha-2E_\beta+E_\gamma))\right],\nonumber
\ea
where $R_3(E_\alpha,E_\beta,E_\gamma)$ is the three-point spectral correlation function~\cite{MehtaBook}. It can be written as
\ba
R_3(E_\alpha,E_\beta,E_\gamma)&=&R_1(E_\alpha)R_1(E_\beta)R_1(E_\gamma)-R_1(E_\alpha)T_2(E_\beta,E_\gamma)\nonumber\\
&-&R_1(E_\beta)T_2(E_\alpha,E_\gamma)-R_1(E_\gamma)T_2(E_\alpha,E_\beta)\nonumber\\
&+&T_3(E_\alpha,E_\beta,E_\gamma).
\ea
In the above,
\be
R_1(E)=\frac{1}{\pi}\sqrt{2D-E^2}
\ee
is the density of states, while $T_{2}$ and $T_{3}$ are the two- and three-point cluster functions, respectively. $T_3$ represents the connected part of the three-point correlation function, while all other terms of $R_3$ represent all possible disconnected contributions.

We first compute the term which depends on the density of states only,
\begin{widetext}
\ba
\frac{(D-3)!}{D!}\int dE e^{-iEt} &\times&\int dE_\alpha dE_\beta dE_\gamma R_1(E_\alpha)R_1(E_\beta)R_1(E_\gamma)\left[\delta(E-(2E_\alpha-E_\beta-E_\gamma))+\delta(E-(E_\alpha-2E_\beta+E_\gamma))\right]\nonumber\\
&=&\frac{(D-3)!}{D!}\int dE_\alpha e^{-2iE_\alpha t}R_1(E_\alpha)\int dE_\beta e^{-iE_\beta t}R_1(E_\beta)\int dE_\gamma e^{-iE_\gamma t}R_1(E_\gamma)+\textrm{c.c.}\nonumber\\
&=&2\frac{(D-3)!}{D!}(2D)^3b_1(2\Gamma t)b_1^2(\Gamma t).
\ea
\end{widetext}

We now consider the terms containing both $R_1$ and $T_2$. As an example, let us compute
\begin{widetext}
\ba
\frac{(D-3)!}{D!}\int dE e^{-iEt} &\times&\int dE_\alpha dE_\beta dE_\gamma R_1(E_\alpha)T_2(E_\beta,E_\gamma)\left[\delta(E-(2E_\alpha-E_\beta-E_\gamma))+\delta(E-(E_\alpha-2E_\beta+E_\gamma))\right]\nonumber\\
&=&\frac{(D-3)!}{D!}\int dE_\alpha R_1(E_\alpha)e^{-2iE_\alpha t}\int dE_\beta dE_\gamma T_2(E_\beta,E_\gamma)e^{-i(-E_\beta-E_\gamma)t}\label{eq:R1T2}\\
&+&\frac{(D-3)!}{D!}\int dE_\alpha R_1(E_\alpha)e^{-iE_\alpha t}\int dE_\beta dE_\gamma T_2(E_\beta,E_\gamma)e^{-i(-2E_\beta+E_\gamma)t}.\nonumber
\ea
\end{widetext}
From Ref.~\cite{MehtaBook}, we know that, for any $n\geq2$, integrals of the form
\be
\int dE_1dE_{2\cdots} dE_nT_n(E_1,E_2,_{\cdots} E_n)e^{-i\sum_{j=1}^n k_j E_j}
\ee
are non-vanishing if and only if $\sum_{j=1}^n k_j=0$. Since this is not the case in the integrals of Eq.~(\ref{eq:R1T2}), all terms in $g_3$ containing both $R_1$ and $T_2$ vanish.

Finally, we consider the term containing the $T_3$ connected correlation function,
\begin{widetext}
\ba
\frac{(D-3)!}{D!}\int dE e^{-iEt} &\times&\int dE_\alpha dE_\beta dE_\gamma T_3(E_\alpha,E_\beta,E_\gamma)\left[\delta(E-(2E_\alpha-E_\beta-E_\gamma))+\delta(E-(E_\alpha-2E_\beta+E_\gamma))\right]\nonumber\\
&=&\frac{(D-3)!}{D!}\int dE_\alpha dE_\beta dE_\gamma T_3(E_\alpha,E_\beta,E_\gamma)\left[e^{-i(2E_\alpha-E_\beta-E_\gamma)t}+e^{-i(E_\alpha-2E_\beta+E_\gamma)t}\right].
\ea
\end{widetext}
We perform the change of variables $\xi_\alpha=E_\alpha/R_1(0)$, and similarly for $E_\beta$ and $E_\gamma$. This corresponds to rescale the energies by their mean level spacing. Calling $Y_3(\xi_\alpha,\xi_\beta,\xi_\gamma)=R_1^3(0)T_3(E_\alpha,E_\beta,E_\gamma)$, this integral becomes
\ba
\frac{(D-3)!}{D!}\int d\xi_\alpha d\xi_\beta d\xi_\gamma Y_3(\xi_\alpha,\xi_\beta,\xi_\gamma)&\times&\nonumber\\
\left[e^{-i(2\xi_\alpha-\xi_\beta-\xi_\gamma)R_1(0)t}+e^{i(\xi_\alpha-2\xi_\beta+\xi_\gamma)R_1(0)t}\right] ,
\label{eq:b3}
\ea
which can be computed using the following formula, found in Ref.~\cite{MehtaBook},
\ba
&&\int d\xi_1d\xi_{2\cdots} d\xi_nY_n(\xi_1,\xi_2,_{\cdots},\xi_n)e^{-i\sum_{j=1}^n\xi_j\tau_j} \nonumber \\
&=&\delta\left(\tau_1+\tau_2+_{\cdots}+\tau_n\right)
\int_{-\infty}^\infty d\tau   \left[  \sum_Pf(\tau)f(\tau+\tau_{P(1)})_{\cdots}  \right.
\nonumber \\
&& f(\tau+\tau_{P(1)}+ _{\cdots}+  \left. \tau_{P(n-1)})   \right]_{(0)}.\nonumber
\label{eq:P}
\ea
In this formula, $P$ labels all permutations of the indexes $1,2,3, \ldots n$, and $P(i)$ is the permuted counterpart of $i$, according to $P$. $f(\tau)$ is a matrix valued function, which reads
\be
f(\tau)=\begin{pmatrix}
f_2(\tau) & \tau f_2(\tau) \\
\frac{f_2(\tau)-1}{\tau} & f_2(\tau)
\end{pmatrix},
\quad
f_2(\tau)=\begin{cases}
1 & \left|\tau\right|<1/2 \\
0 & \left|\tau\right|\geq1/2
\end{cases}.
\ee
With the notation $\left[\cdot\right]_{(0)}$, we mean the following: Taken a generic $2\times2$ matrix $A$, we call
\be
\left[A\right]_{(0)}=\frac{A_{11}+A_{22}}{2},
\ee
where $A_{11}$ and $A_{22}$ are the diagonal entries of $A$. In particular, for Eq.~(\ref{eq:b3}), we have $\tau_1=2R_1(0)$, $\tau_2=\tau_3=-R_1(0)$ for the first term, and $\tau_1=\tau_3=R_1(0)$, $\tau_2=-2R_1(0)$ for the second. Thus, we obtain for Eq.~(\ref{eq:b3}), 
\ba
&&\frac{(D-3)!}{D!}\int d\xi_\alpha d\xi_\beta d\xi_\gamma Y_3(\xi_\alpha,\xi_\beta,\xi_\gamma) \nonumber \\
&& \times \left[e^{-i(\xi_\alpha-\xi_\beta-\xi_\gamma)R_1(0)t}+e^{i(\xi_\alpha-2\xi_\beta+\xi_\gamma)R_1(0)t}\right] \nonumber \\
&=&\frac{(D-3)!}{D!}Db_3\left(\frac{\Gamma t}{2D}\right),
\ea
with
\be
b_3(\tau)=\begin{cases}
4\left[1-4\tau+3\tau\log\left(2\tau+1\right)\right] & 0<\tau\leq\frac{1}{2}\\
8\left[\tau-1+\frac{3}{2}\tau\log\left(\frac{4-\tau}{2+\tau}\right)\right] & \tau>\frac{1}{2}
\end{cases}.
\ee

Putting all these terms together, we get
\be
g_3(t)=\frac{(D-3)!}{D!}\left[16D^3b_1(2\Gamma t)b_1^2(\Gamma t)-Db_3\left(\frac{\Gamma t}{2D}\right)\right].
\ee

\subsection{Term $\pmb{\Xi_1}$}

Finally, we compute the term $\Xi_1$,
\be
\left<\sum_{\alpha\neq\gamma\neq\beta\neq\delta}e^{-i(E_\alpha-E_\beta+E_\gamma-E_\delta)t}\left|c^{(0)}_\alpha\right|^2\left|c^{(0)}_\beta\right|^2\left|c^{(0)}_\gamma\right|^2\left|c^{(0)}_\delta\right|^2\right>.
\ee
Using again the independence of eigenvalues and eigenvectors, we factorize the average, and compute the part depending on the eigenvectors first,
\ba
\left<\sum_{\alpha\neq\gamma\neq\beta\neq\delta}\left|c^{(0)}_\alpha\right|^2\left|c^{(0)}_\beta\right|^2\left|c^{(0)}_\gamma\right|^2\left|c^{(0)}_\delta\right|^2\right>&=&\nonumber\\
1-6\left<\overline{P_S}\right>+O(D^{-2})&\sim&1.
\ea

We now compute the average over the energy levels,
\ba
g_4(t)&=&\left<e^{-i(E_\alpha-E_\beta+E_\gamma-E_\delta)}\right>\nonumber\\
&=&\frac{(D-4)!}{D!}\int dE e^{-iEt}\nonumber\\
&\times&\int dE_\alpha dE_\beta dE_\gamma dE_\delta R_4(E_\alpha,E_\beta,E_\gamma,E_\delta)\nonumber\\
&\times&\delta(E-(E_\alpha-E_\beta+E_\gamma-E_\delta)).
\ea
$R_4(E_\alpha,E_\beta,E_\gamma,E_\delta)$ is the four-point correlation function, which can be written as~\cite{MehtaBook}
\begin{widetext}
\ba
R_4(E_\alpha,E_\beta,E_\gamma,E_\delta)&=&R_1(E_\alpha)R_1(E_\beta)R_1(E_\gamma)R_1(E_\delta)-R_1(E_\alpha)R_1(E_\beta)T_2(E_\gamma,E_\delta)-R_1(E_\alpha)R_1(E_\gamma)T_2(E_\beta,E_\delta)\nonumber\\
&-&R_1(E_\alpha)R_1(E_\delta)T_2(E_\beta,E_\gamma)-R_1(E_\beta)R_1(E_\gamma)T_2(E_\alpha,E_\delta)-R_1(E_\beta)R_1(E_\delta)T_2(E_\alpha,E_\gamma)\nonumber\\
&-&R_1(E_\gamma)R_1(E_\delta)T_2(E_\alpha,E_\beta)+T_2(E_\alpha,E_\beta)T_2(E_\gamma,E_\delta)+T_2(E_\alpha,E_\gamma)T_2(E_\beta,E_\delta)\nonumber\\
&+&T_2(E_1,E_4)T_2(E_2,E_3)+R_1(E_1)T_3(E_2,E_3,E_4)+R_1(E_2)T_3(E_1,E_3,E_4)\nonumber\\
&+&R_1(E_3)T_3(E_1,E_2,E_4)+R_1(E_4)T_3(E_1,E_2,E_3)-T_4(E_\alpha,E_\beta,E_\gamma,E_\delta).
\ea
\end{widetext}
Once again, this amounts to consider the connected correlation function $T_4(E_\alpha,E_\beta,E_\gamma,E_\delta)$ and all disconnected components. We now study the resulting integrals, one by one.

We first consider the term containing $R_1$ only:
\begin{widetext}
\ba
\frac{(D-4)!}{D!}&\times&\int dE_\alpha dE_\beta dE_\gamma dE_\delta R_1(E_\alpha)R_1(E_\beta)R_1(E_\gamma)R_1(E_\delta)e^{-i(E_\alpha-E_\beta+E_\gamma-E_\delta)t}\nonumber\\
&=&\frac{(D-4)!}{D!}\left|\int dE R_1(E)e^{-iEt}\right|^4=\frac{(D-4)!}{D!}(2D)^4b_1^4(\Gamma t).
\ea
\end{widetext}

Next, we consider terms containing both $R_1$ and $T_2$, such as
\begin{widetext}
\begin{equation}
\frac{(D-4)!}{D!}\int dE_\alpha dE_\beta dE_\gamma dE_\delta R_1(E_\alpha) R_1(E_\beta) T_2(E_\gamma,E_\delta) e^{-i(E_\alpha-E_\beta+E_\gamma-E_\delta)t}.
\end{equation}
\end{widetext}
This term can be rewritten as
\ba
\frac{(D-4)!}{D!}&\times&\left|\int dE R_1(E)e^{-iEt}\right|^2\\
&\times&\int dE_\gamma dE_\delta T_2(E_\gamma,E_\delta)e^{-i(E_\alpha-E_\beta)t}\nonumber\\
&=&\frac{(D-4)!}{D!}(2D)^2b_1^2(\Gamma t)Db_2\left(\frac{\Gamma t}{2D}\right).\nonumber
\ea
Note that this integral is non zero only if the two energies in the $T_2$ function come with opposite signs in the corresponding exponential. For this reason, this term comes with a combinatorial factor equal to $4$.

Next are the terms containing $T_2$ only,
\begin{widetext}
\ba
\frac{(D-4)!}{D!}\int dE_\alpha dE_\beta dE_\gamma dE_\delta T_2(E_\alpha,E_\beta)T_2(E_\gamma,E_\delta)e^{-i(E_\alpha-E_\beta+E_\gamma-E_\delta)t}&=&\nonumber\\
\frac{(D-4)!}{D!}\left(\int dE_\alpha dE_\beta T_2(E_\alpha,E_\beta)e^{-i(E_\alpha-E_\beta)t}\right)^2&=&\frac{(D-4)!}{D!}D^2b_2^2(\Gamma t).
\ea
\end{widetext}
This term comes with a combinatorial factor equal to $2$.

Then come the terms with $R_1$ and $T_3$, such as
\ba
\frac{(D-4)!}{D!}&\times&\int dE_\alpha dE_\beta dE_\gamma dE_\delta\\
&\times&R_1(E_\alpha)T_3(E_\beta,E_\gamma,E_\delta)e^{-i(E_\alpha-E_\beta+E_\gamma-E_\delta)t}.\nonumber
\ea
However, as explained for the $g_3(t)$ function, all the integrals of this form vanish. 

Finally, comes the term containing the $T_4$ function:
\ba
\frac{(D-4)!}{D!}&\times&\int dE_\alpha dE_\beta dE_\gamma dE_\delta T_4(E_\alpha,E_\beta,E_\gamma,E_\delta)\\
&\times&e^{-i(E_\alpha-E_\beta+E_\gamma-E_\delta)t}=\frac{(D-4)!}{D!}Db_4\left(\frac{\Gamma t}{2D}\right).\nonumber
\ea
We compute the function $b_4$ with a procedure analogous to the one used for $b_3$. The result is
\begin{widetext}
\be
b_4(\tau)=\begin{cases}
\frac{4\tau^3-20\tau^2-4\tau+3}{2\tau+1}+6\tau\log(2\tau+1) & \tau\leq\frac{1}{2} \\
2\frac{2\tau^3+8\tau^2-2\tau-3}{2\tau+1}+3\tau\left[2\log(2\tau+1)-3\log(4\tau-1)+\log2\right] & \frac{1}{2}<\tau<1 \\
2\frac{6\tau^2-1}{4\tau^2-1}+3\tau\log\left(\frac{2\tau+1}{2\tau-1}\right)
\end{cases}.
\ee
\end{widetext}
So the $g_4$ function reads
\begin{widetext}
\be
g_4(t)=\frac{(D-4)!}{D!}\left[16D^4b_1^4(\Gamma t)-16D^3b_1^2(\Gamma t)b_2\left(\frac{\Gamma t}{2D}\right)+2D^2b_2^2\left(\frac{\Gamma t}{2D}\right)-Db_4\left(\frac{\Gamma t}{2D}\right)\right].
\ee
\end{widetext}
Combining all the terms together, one recovers Eq.~(\ref{eq:PS2analytic}) from the main text.

\section{Plots for the relative variance of the second-order R\'enyi entropy} 
\label{app:Rs}

\begin{figure}[ht]
\includegraphics*[width=0.49\textwidth]{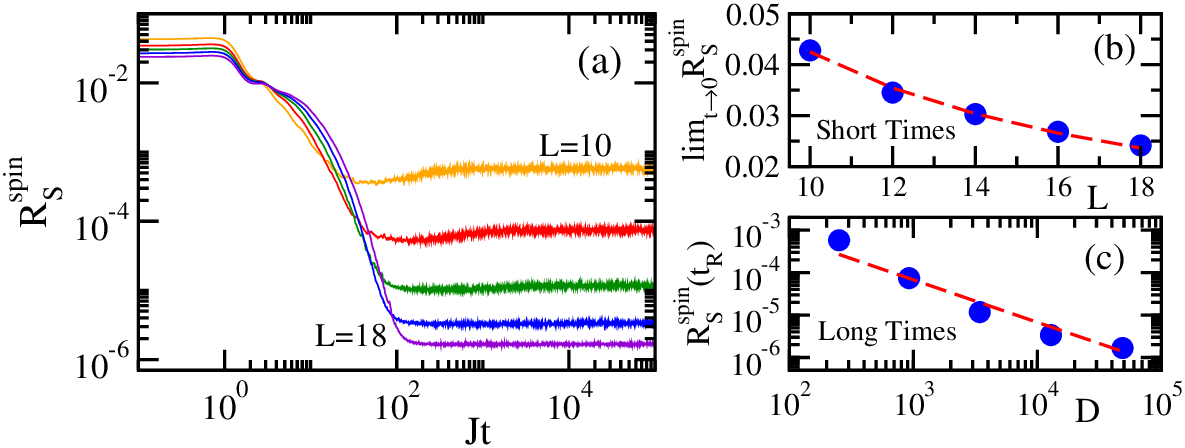}
\caption{Relative variance of the second-order R\'enyi entropy for the spin model (a). The system sizes are indicated in (a): From the top curve to the bottom curve (at large times), the sizes of the matrices are $D=252,924,3\,432,12\,870,48\,620$ (orange, red, green, blue, purple). In (b): coefficient $\lim_{t\rightarrow0}{\cal R}_{S}^{\text{spin}}(t)/t^4$  as a function of $L$; numerical values (circles) and theoretical estimate $\propto 1/L$ (dashed line). In (c): numerical values (circles) of ${\cal R}_{S}^{\text{spin}} (t)$ for $t>t_{\text{R}}$ as a function of $D$ and curve $\propto 1/D$ (dashed lines).}
\label{fig:S}
\end{figure}

For completeness, we present in Fig.~\ref{fig:S}~(a) the relative variance of the second-order R\'enyi entropy as a function of time. The behavior for the relative variance of the Shannon entropy is very similar (not shown). 

Contrary to the inverse participation ratio, the second-order R\'enyi entropy is self-averaging at short times. Indeed, as seen in Fig.~\ref{fig:S}~(b), the relative variance follows very well the prediction in Eq.~(\ref{Eq:SatShortTime}) that  ${\cal R}_S^\textrm{spin}(t) \propto 1/L$, as we find also for the spatially local quantities discussed in Sec.~\ref{Sec:Imb}. 

At long times, the second-order R\'enyi entropy behaves similarly to the inverse participation ratio and the connected spin-spin correlation function, being super self-averaging, since ${\cal R}_S^\textrm{spin}(t) \propto 1/D$, as shown in Fig.~\ref{fig:S}~(c).  The behavior of ${\cal R}_S^\textrm{spin}(t)$ for short- and long-times is therefore similar to that of ${\cal R}_C^\textrm{spin}(t)$.

Between the two extremes of short and long times, in the region of  the power-law behavior of $\langle P_S (t) \rangle$, the second-order R\'enyi entropy is not self-averaging. This means that the relative variance of $S$ (same for $Sh$) exhibits two crossing points as time increases from zero, a feature that contrasts with those of the other quantities studied in this paper, $P_S$, $\text{IPR}$, $I$, and $C$.

We fitted ${\cal R}_{S}^{\text{spin}} (t>t_{\text{R}})$ in Fig.~\ref{fig:S}~(c) and ${\cal R}_{\text{IPR}}^{\text{spin}} (t>t_{\text{R}})$ in Fig.~\ref{fig:IPR}~(f) with $1/D$ for making an analogy with the results for the GOE model, but the exponent $\nu$ in the best fit $1/D^\nu$ is $1\pm 0.2$. In Fig.~\ref{fig:I}~(f) for the connected spin-spin correlation function, where we do not make a comparison with the GOE model, we actually show the best fit. Independently of the exact value of the exponent $\nu$, the relative variance at long times for any of these three quantities decreases exponentially fast with $L$.



%

\end{document}